\documentclass[a4paper,11pt]{article}
\usepackage[british]{babel}

\usepackage[includeheadfoot, width=432pt, height=720pt]{geometry}
\usepackage[utf8]{inputenc}
\usepackage{hyperref}
\hypersetup{colorlinks=false, linkcolor=red, citecolor=green, urlcolor=blue-purple}
 \usepackage[splitrule]{footmisc} 
 
\usepackage{changepage}

\usepackage{booktabs}
\usepackage{siunitx}
\usepackage{verbatim}
\usepackage{lmodern}
\usepackage{lipsum}
\usepackage{booktabs}
\usepackage{caption}
\usepackage{cite}
\usepackage{soul,color}
\usepackage{titlesec}
\usepackage{tabularx}
\usepackage{graphicx}
\usepackage[table]{xcolor}
\usepackage[autostyle]{csquotes}
\usepackage{amsthm, amssymb,amsfonts}
\usepackage{amsmath}
\usepackage{array}
\numberwithin{equation}{section}
\usepackage{bm}
\usepackage{mathtools}
\usepackage{amstext}
\usepackage{braket}
\usepackage{multirow}
\usepackage[normalem]{ulem}

\newcommand{\nc}{\newcommand}
\nc{\lb}{\llbracket}
\nc{\rb}{\rrbracket}
\nc{\gl}{\llbracket}
\nc{\gr}{\rrbracket}

\def\be{\begin{equation}}
\def\ee{\end{equation}}
\def\beq{\begin{eqnarray}}
\def\eeq{\end{eqnarray}}
\nc{\gm}{$m_{3/2}$ }

\newcommand{\bal}{\begin{aligned}}   
\newcommand{\eal}{\end{aligned}}
\newcommand{\bea}{\begin{eqnarray}}  
\newcommand{\eea}{\end{eqnarray}}

\newcolumntype{C}[1]{>{\centering\arraybackslash}m{#1}}

\begin{document}

\begin{titlepage}

\vspace*{-2cm} 
\begin{flushright}
{\tt \phantom{xxx}  MPP-2026-107} \qquad \qquad
\end{flushright}

\vspace*{0.8cm} 
\begin{center}
{\Huge Supersymmetry, Large Extra
\vspace*{0.5cm}
Dimensions and the Gravitino Conjecture
}\\

 \vspace*{1.5cm}
{\bf Leonardo Bersigotti}$^{1}$, {\bf Dieter Lüst}$^{1,\,2}$ and {\bf Marco Scalisi}$^{1,\,3,\,4}$ \\

\begin{flushleft}
\begin{small}
 \vspace*{0.7cm} 
$^1$ {\it Max-Planck-Institut f\"ur Physik, Boltzmannstrasse 8, 85748 Garching bei München, Germany}\\[2mm]

$^2$ {\it Arnold Sommerfeld Center for Theoretical Physics, Ludwig-Maximilians-Universit\"at M\"unchen, 80333 M\"unchen, Germany}\\[2mm]

$^3$ {\it Department of Physics and Astronomy “Ettore Majorana”, University of Catania, Via S. Sofia 64, 1-95125 Catania, Italy
}\\[2mm]

$^4$ {\it INFN-Sezione di Catania, Via Santa Sofia 64, I-95123 Catania, Italy
}\\[2mm]
\end{small}
\end{flushleft}

\begin{abstract}

\noindent
We investigate whether the absence of experimental signals for supersymmetry and extra dimensions can be understood as a correlated phenomenon. Assuming the Gravitino Conjecture, we study the relation between the gravitino mass and the Kaluza--Klein scale in four-dimensional \(\mathcal N=1\) supergravity from Type II compactifications with large extra dimensions. We parametrize the scaling of the full internal volume with respect to the one of a large \(p\)-cycle through an anisotropy exponent \(\alpha\), and derive the corresponding volume contributions to the K\"ahler potential. This leads to constraints on the scaling exponent \(n\), linking the gravitino mass to the KK scale, and to the effective number \(\alpha p\) of large dimensions. We find that the linear relation \(n=1\) is compatible only with one or two large extra dimensions, precisely the cases that can still be probed at micron distances. In such scenarios, micron-sized extra dimensions imply a light gravitino and gauge-mediated supersymmetry breaking, whereas gravity mediation corresponds to compactification scales beyond current experimental reach.

\end{abstract}
\end{center}
\vspace{2cm}
\end{titlepage}

\tableofcontents 
\newpage

\section{Introduction}

Supersymmetry (SUSY) and extra dimensions play pivotal roles in several prominent proposals that extend beyond our current understanding of particle physics and gravity. Yet, there is no experimental evidence for either of them.

The bounds set by the Large Hadron Collider (LHC) on the mass of the heaviest supersymmetric particle \cite{ParticleDataGroup:2024cfk} (of order TeV) automatically translate into a universal lower bound on the mass of the gravitino\footnote{Given the extremely small observed dark energy density, the gravitino mass can be regarded as an order parameter of SUSY breaking, through the relation $M^2_{\text{SUSY}} \simeq m_{3/2}\ M_{\text{P}}$, where $M_{\text{SUSY}}$ and $M_{\text{P}}$ denote the SUSY breaking scale and the reduced Planck mass, respectively.} $m_{3/2}\geq \mathcal{O}(0.1)\ {\rm eV}$. This can be equivalently expressed as a universal upper bound on its associated Compton wavelength,
\begin{equation}
    r_{3/2}\leq \mathcal{O}({\rm \mu m})\ \,.
\end{equation}
Intriguingly, this scale coincides with the order of magnitude above which no experimental evidence for extra dimensions has ever been observed. Astrophysical analyses of supernova explosions~\cite{Hannestad:2004aa} and precision tests of Newton’s law \cite{Lee:2020aa} suggest that one or even two {\it large} extra dimensions\footnote{We call `large' those extra dimensions \textit{compactified} at scales that are large enough to potentially affect low-energy physics, yet small enough to remain consistent with current experimental bounds.} are not excluded, provided their size does not exceed the micrometer range.

This coincidence in the corresponding upper bounds encourages us to speculate whether the two phenomena might be connected, and then whether the absence of observed deviations from Newton’s law is in any way related to the non-observation of SUSY.

In the context of the Swampland program \cite{vafa2005string,Ooguri_2007,Palti:2019aa,van_Beest_2022}, it has been conjectured that the gravitino mass is intrinsically linked to the mass scale of an infinite tower of states. In particular, the massless limit $m_{3/2}\rightarrow 0$ is associated with a lowering of the quantum-gravity cutoff and the consequent breakdown of the effective field theory (EFT). This proposal is referred to in the original works as the Gravitino Mass Conjecture \cite{Cribiori:2021aa} and the Gravitino Distance Conjecture \cite{Castellano:2021aa}; throughout this work, we will collectively refer to it as the {\it Gravitino Conjecture} (GC). In the limits considered in this work, the tower of states is identified with Kaluza--Klein (KK) modes and, in the simplest scenario, it is possible  to relate the KK mass scale   $m_{\text{KK}}$   and the gravitino mass as
\begin{equation}
\label{firstGC}
    m_{\text{KK}} \sim (m_{3/2})^n\,,
\end{equation}
with $n$ being an $\mathcal{O}(1)$ parameter. Expressed in terms of the associated length scales, the relation becomes
\begin{equation}\label{}
    R \sim (r_{3/2})^{n}\,,
\end{equation}
with $R=(m_{\text{KK}})^{-1}$ being the typical size of the (large) extra dimensions. Even at this schematic level, the equation above provides a natural framework to interpret the coincidence of bounds discussed above, particularly when $n=1$.

Large extra dimensions have been considered in the literature to address long-standing physics puzzles, such as the electroweak hierarchy problem, as proposed in \cite{Arkani-Hamed:1998jmv, Antoniadis_1998, Arkani_Hamed_1999, Arkani_Hamed_2001}, and the smallness of the cosmological constant, as argued in the more recent Dark Dimension (DD) scenario \cite{Montero:2022prj}. In particular, the DD proposal predicts one or possibly two \cite{Anchordoqui:2025nmb}
extra dimensions with a size precisely in the micron range and relates this scale to the extremely small value of the dark energy density. This intriguing connection has been further investigated in the context of cosmology and particle physics in several works \cite{Anchordoqui:2022txe,Gonzalo:2022jac,Anchordoqui:2022tgp,Anchordoqui:2022svl,Anchordoqui:2023oqm,Noble:2023mfw,Anchordoqui:2023tln,Law-Smith:2023czn,Obied:2023clp,Anchordoqui:2024akj,Gendler:2024gdo,Heckman:2024trz,Basile:2024lcz,Anchordoqui:2025xug,Anchordoqui:2025epz,Eller:2025lsh}.

In Ref.~\cite{Anchordoqui:2023oqm}, exploiting the Gravitino Conjecture, a relation between the value of the dark energy density and the SUSY-breaking scale was proposed, leading to the striking prediction that signatures of supersymmetry might appear at energies in the range $\mathcal{O}(10\text{--}100)$ TeV. Despite an apparent similarity, this line of investigation is conceptually distinct from the present work, since the value of the dark energy density plays no direct role in our current analysis. In particular, we do not necessarily require its magnitude to be connected to the mass scale of an infinite tower of states, as it would be implied by the Anti-de Sitter Distance Conjecture (ADC) \cite{Lust:2019zwm}.

In this work, we investigate the phenomenological implications of the direct relation between the gravitino mass and the Kaluza--Klein scale, as given in eq.~\eqref{firstGC}. Furthermore, in a broad class of string compactifications, large extra dimensions are often realized through cycles whose directions are parametrically larger than the remaining ones, although this does not necessarily imply that their volume dominates the total internal volume. We study the situation in general terms, assuming that $p$ dimensions become large and by parametrizing the scaling of the full internal volume with respect to
the corresponding \(p\)-cycle volume, in string units, as
\begin{equation}
    \mathcal V \sim \mathcal V_p^\alpha\,.
\end{equation}
The exponent \(\alpha\) encodes the anisotropy of the internal space.
Within this setting, we analyse $\mathcal{N}=1$ supergravity descriptions of Type IIA and Type IIB compactifications and show that the volume-dependent part of the K\"ahler potential becomes
\begin{equation}
K^{\mathcal{V}}_{IIA}= - \alpha\,\ln(\mathcal{V}_p)\,,\qquad  
K^{\mathcal{V}}_{IIB}= - 2\alpha\,\ln(\mathcal{V}_p)\,.
\end{equation}
which reproduces the standard expressions for the overall volume by replacing $\mathcal{V}\sim\mathcal{V}^\alpha_p$.

This result allows us to derive sharp constraints on the exponent $n$ appearing in eq.~\eqref{firstGC} as a function of the number $p$ of large extra dimensions. Intriguingly, we find that $n=1$, which without severe fine-tuning would account for the coincidence of bounds at the micrometer scale discussed above, is excluded for any $p \geq 3$. This theoretical outcome is in remarkable agreement with current experimental constraints, which single out scenarios with one or two large extra dimensions as the only cases still compatible with the micron scale.
Taken together, these findings allow us to present a complete classification of physical scenarios consistent with current particle physics and astrophysical bounds. Given the constraints on the parameter $n$, we show that extra dimensions in the micron range correspond to low gravitino masses in the eV–GeV range, implying that supersymmetry breaking must proceed via gauge mediation. Alternatively, scenarios with a much heavier gravitino, $m_{3/2}\sim\mathrm{TeV}$, corresponding to gravity-mediated supersymmetry breaking, are associated with extra dimensions whose size lies beyond current experimental sensitivity.

The structure of this work is as follows. In Section~\ref{gravitino:conj:section}, we
review the Gravitino Conjecture (GC) and fix our conventions for four-dimensional
\(\mathcal{N}=1\) supergravity. Section~\ref{sec:bounds:on:extra:dimensions} summarizes
the relevant experimental and astrophysical bounds on \(m_{3/2}\), \(m_{\rm KK}\), and the
number of large extra dimensions. Section~\ref{section:compactification:generalities}
introduces Type IIA and Type IIB compactifications with an internal volume expressed in
terms of a large \(p\)-cycle, derives the anisotropy-dependent K\"ahler-potential
prefactors, and discusses the role of the parameter \(\alpha\). These results are used in
Section~\ref{sec:volume:relations} to obtain the perturbative bounds on the
Gravitino-Conjecture exponent \(n\), emphasizing that, at fixed values of $\alpha$ correspond an interval structure in Type IIA and
a fixed value for $n$ in Type IIB. Phenomenological implications are discussed in Section~\ref{chap:finding:scenarios}, where we analyze the allowed scenarios for $\Lambda_{\rm sp}$, $M_{\text{SUSY}}$, $m_{\text{KK}}$ and \gm as functions of $p$ and $\alpha$. Finally, three appendices complete the paper collecting conventions and technical details. In Appendix~\ref{AppII} we specify our conventions for Type II Calabi--Yau orientifold compactifications, in Appendix~\ref{sec:m_kk:compactification} we summarize the Kaluza--Klein and dimensional-reduction conventions, while
Appendix~\ref{app:species:scale:anisotropy} contains the species-scale analysis in anisotropic compactifications.

\section{Gravitino Conjecture and \texorpdfstring{$\mathcal{N}=1$}{N=1} SUGRA}\label{gravitino:conj:section}

The GC, as proposed in both original works~\cite{Cribiori:2021aa,Castellano:2021aa}, states the following:\footnote{In this work we will always refer to the Lagrangian mass of the gravitino.}

\begin{adjustwidth}{1.8em}{1.8em}
\textit{In a theory with non-vanishing gravitino mass $m_{3/2}$, the limit $m_{3/2} \rightarrow 0$ always corresponds to the massless limit of an infinite tower of states and to the breakdown of the EFT}.
\end{adjustwidth}
According to this, vacua with $m_{3/2}=0$ are not in the Swampland; rather, the conjecture states that \gm should not be allowed to vary continuously towards zero. For our purposes, the value of \gm may also be far from zero, as long as it remains small in Planck units and tends to zero only in the asymptotic limits of the EFT. In string theory, however, \gm is moduli-dependent. Therefore, taking $m_{3/2}\rightarrow0$ corresponds to moving toward asymptotic regions of moduli space.

The appearance of a massless infinite tower of states in the Gravitino Conjecture (GC) should be understood in close analogy with the Swampland Distance Conjecture (DC) \cite{Ooguri_2007}. The states in the tower can correspond either to Kaluza--Klein (KK) modes or to excitations of a single string \cite{Lee:2019wij}. In this work, we focus on the case in which the limit $m_{3/2}\to 0$ corresponds to a decompactification limit of perturbative string theory, relying on a specific behavior of the string coupling, as it will be discussed in Section~\ref{sec:perturbative:regime:string:theory}.

The scaling relation between the KK mass scale and the gravitino mass, proposed in \cite{Cribiori:2021aa,Castellano:2021aa}, takes the form of eq.~\eqref{firstGC} and can be written explicitly as
\begin{equation}
\label{GravConjFinal}
m_{\text{KK}}=\lambda_{3/2}\left(\frac{m_{3/2}}{M_{\text{P}}}\right)^n M_{\text{P}}\,,
\end{equation}
where the parameter $\lambda_{3/2}$ is assumed to be equal to unity throughout this work.

The validity of this relation is tested in several examples of string theory in Minkowski and anti-de Sitter backgrounds. In the anti-de Sitter case, the supersymmetric contribution of $m_{3/2}$ is likewise expected to realize the conjecture, making the statement \eqref{GravConjFinal} equivalent to the anti-de
Sitter distance conjecture \cite{Lust:2019zwm}. In \cite{Cribiori:2021aa}, it is suggested that the GC holds also in de Sitter spacetimes. 
Furthermore, we focus on the GC ignoring any log corrections \cite{blumenhagen_quantum_2020,Cribiori:2021aa}. Moreover, the existence of an infinite tower of light states associated to fermions in quantum gravity was already suggested in \cite{Palti:2020tsy}.

The low-energy SUGRA description, throughout this work, is described by the four-dimensional \mbox{$\mathcal{N}=1$} SUGRA action coupled to matter \cite{freedman_supergravity_2013}. The low-energy degrees of freedom consist of the gravity multiplet, containing the spin-2 graviton $g_{\mu \nu}$ and the spin-3/2 gravitino $\Psi_{\mu}$. In addition, the theory contains $n_C$ chiral multiplets $\Phi^i=(\phi^i, \psi^i)$, where $\phi^i$ are complex scalar fields parametrizing the Kähler manifold of the theory, and $\psi_i$ are their spin-1/2 Weyl fermionic partners. The theory also includes $n_V$ vector multiplets $V^a=(\lambda^a, A^a_{\mu})$, where $A^a_{\mu}$ are the spin-1 gauge fields associated with the gauged isometries of the scalar manifold and $\lambda^a$ are their spin-1/2 gaugino partners.
The full structure of four-dimensional $\mathcal{N}=1$ supergravity is completely determined by the Kähler potential $K(\phi^i, \bar{\phi}^{\bar{\jmath}})$, the holomorphic superpotential $W(\phi^i)$,
the gauge kinetic function $f_{ab}(\phi^i)$, and the Killing vectors $k_a = k_a^{\, i}(\phi)\, \partial / \partial \phi^i$ that generate the isometries of the scalar manifold gauged by the vector fields \cite{DallAgata:2021uvl}. The gravitino mass in Planck units is then
\begin{equation}
    \label{field:dependent::gravitino:mass}
    m_{3/2} = e^{K(\phi,\bar{\phi})/2} W(\phi)\,,
\end{equation}
and the scalar potential is
\begin{equation}
    \label{scalar:potential}
    V = e^{K} \left( K^{i\bar{j}} D_i W\, D_{\bar{j}} \bar{W} - 3 |W|^2 \right)
        + \frac{1}{2} (\mathrm{Re}\, f)^{-1\,ab} D_a D_b\,,
\end{equation}
where $D_i W = \partial_i W + (\partial_i K)\, W$ is the Kähler-covariant derivative, and 
$K^{i\bar{j}}$ is the inverse Kähler metric, i.e. the inverse of 
$K_{i\bar{j}} = \partial_i \partial_{\bar{j}} K$. The first two terms in \eqref{scalar:potential} contain the positive F-term contribution $V_F$ and the negative gravitational part. The last term is the D-term contribution with
$D_a = i K_j k^j_a + \xi_a$, where $\xi_a$ are the Fayet--Iliopoulos terms associated with the gauging of a $U(1)$ $R$-symmetry. Note that pure D-term supersymmetry breaking is not possible unless $W=0$. 
However, in what follows we will restrict to pure F-term supersymmetry breaking and neglect D-terms for simplicity.

Having defined the gravitino mass as in \eqref{field:dependent::gravitino:mass}, we can rewrite the scalar potential as
\begin{equation}
\label{scalar:potential:with:gravitino:mass}
    V = V_F - 3 m_{3/2}^{2} M_{\rm P}^2\,,
\end{equation}
where we have restored explicitly the dependence on the four-dimensional Planck mass $M_{\rm P}$.

Considering pure $F$-term SUSY breaking at a scale $M_{\rm SUSY}$, we can write $V_F \sim M_{\rm SUSY}^4$, so that \eqref{scalar:potential:with:gravitino:mass} becomes
\begin{equation}
    V \simeq M_{\rm SUSY}^4 - 3 m_{3/2}^{2} M_{\rm P}^2\,.
\end{equation}
In the case of a quasi-flat universe, i.e. with vanishing scalar potential $V \simeq 0$, we obtain
\begin{equation}
\label{gravitino:and:SUSY:breaking:scale}
    M_{\rm SUSY}^2 \simeq m_{3/2} M_{\rm P}\,.
\end{equation}

\section{Bounds on extra dimensions and the gravitino mass}\label{sec:bounds:on:extra:dimensions}

In this section, we provide a comprehensive classification of the theoretical and phenomenological bounds on $m_{\rm KK}$ (or equivalently, its inverse $R=m_{\rm KK}^{-1}$) and $m_{3/2}$, highlighting the physical motivations behind them. This section is fundamental for the subsequent analysis of phenomenological scenarios.

\subsection{Bounds on extra dimensions}

Supernova explosions provide a powerful laboratory for probing the existence of extra dimensions~\cite{Arkani_Hamed_1999}. 
In models with large extra dimensions, the cores of supernovae can efficiently produce Kaluza--Klein (KK) gravitons. 
The requirement that old neutron stars are not excessively heated by a cloud of gravitationally trapped KK modes leads to stringent astrophysical constraints on the compactification scale, 
$R = m_{\text{KK}}^{-1}$, derived in Ref.~\cite{Hannestad:2004aa}. 
These bounds translate into upper limits on the allowed size of the $ p $ large extra dimensions and are summarized in table~\ref{tab:astrophisical:bounds:on:l}.

\begin{table}[h]
    \centering
    \begin{tabular}{|c| c|}
        \hline
        $p$ & \textbf{Upper bound on $R$ [m]} \\
        \hline
        1 & $8.3$ \\
        2 & $5.9\times10^{-8}$ \\
        3 & $1.3\times10^{-10}$ \\
        4 & $5.9\times10^{-12}$ \\
        5 & $9.4\times10^{-13}$ \\
        6 & $2.8\times10^{-13}$ \\
        \hline
    \end{tabular}
    \caption{Upper limits on the size $R$ of $p$ large extra dimensions derived from neutron-star constraints.
    These bounds apply when the internal manifold possesses isometries.}
    \label{tab:astrophisical:bounds:on:l}
\end{table}

However, these constraints on neutron-star heating rely on the assumption that KK modes can decay only into Standard Model fields localized on the brane, and cannot decay into other modes with smaller bulk momenta within the same KK tower, i.e., intra-tower decay. This assumption follows from KK momentum conservation in the graviton tower, but it can be relaxed if the extra dimensions do not possess continuous isometries. In such a case, the internal manifold does not admit non-trivial solutions to the Killing equations. Consequently, a given graviton mode in the KK tower can decay into final states that include other, lighter KK graviton excitations, which becomes the dominant decay channel~\cite{Mohapatra_2003}.
As a result, the partial decay width $\Gamma$ must be enlarged to account for a slight violation of the KK quantum number~\cite{Gonzalo:2022jac, Anchordoqui:2025nmb}. By considering this effect and applying a correction factor related to KK momentum violation, Ref.~\cite{Anchordoqui:2025nmb} shows that the bounds reported in table~\ref{tab:astrophisical:bounds:on:l} can be evaded. In this case, the relevant constraints are obtained by requiring that SN 1987A did not emit more KK gravitons than compatible with the observed neutrino signal duration. These bounds, also discussed in Ref.~\cite{Hannestad:2004aa}, are less stringent than the previous ones, allowing for one or two extra dimensions with $R \sim \mu\text{m}$~\cite{Anchordoqui:2025nmb}. table~\ref{tab:correct:astrophisical:bounds:on:l} presents the updated astrophysical constraints.

\begin{table}[h]
    \centering
    \begin{tabular}{|c| c|}
        \hline
        $p$ & \textbf{Upper bound on} $R\,[\mathrm{m}]$ \\
        \hline
         1& $4.9 \times 10^{2}$\\
         2& $9.6 \times 10^{-7}$\\
         3& $1.14 \times 10^{-9}$\\
         4& $3.82 \times 10^{-11}$\\
         5& $4.85 \times 10^{-12}$\\
         6& $1.21 \times 10^{-12}$\\
        \hline
    \end{tabular}
    \caption{Upper limits on the size $R$ of $p$ large extra dimensions derived from neutron-star constraints.
    These bounds apply more generally also when the internal manifold does not possess isometries.}
    \label{tab:correct:astrophisical:bounds:on:l}
\end{table}

From the experimental side, considering large extra dimensions, one should also account for how Newton's gravitational law, $1/r^2$, is expected to be modified due to the presence of KK gravitons. In particular, one expects an additional Yukawa-like contribution to the familiar Newtonian potential. In Ref.~\cite{Lee:2020aa}, the authors did not observe the effects of these additional couplings up to the explored length scale of $38.6 \, \mu\text{m}$. This sets an experimental upper bound on the size of extra dimensions:
\begin{equation}
\label{experimental:bound:on:l}
    R \leq 38.6 \, \mu\text{m}.
\end{equation}
This bound is more stringent only in the case of a single extra dimension ($p=1$), and therefore table~\ref{tab:correct:astrophisical:bounds:on:l} should be updated with this value for $p=1$. 

In summary, only the cases $p=1$ and $p=2$ are compatible with an upper bound of order micrometers for the size of extra dimensions.

\subsection{Bounds on the gravitino mass}
\label{sec:bounds:on:grav:mass}

At present, no direct measurement of the gravitino mass is available, and only bounds can be considered. In a quasi-flat spacetime the relation \eqref{gravitino:and:SUSY:breaking:scale} applies, allowing us to translate the lower bounds on the masses of superpartners set by the LHC in its high-luminosity era, at roughly $\mathcal{O}(1-10)\,\text{TeV}$ \cite{particlereview}, into corresponding bounds on the SUSY-breaking scale $M_{\text{SUSY}}$ and, consequently, on the gravitino mass.

In order to do so, it is necessary to specify the dependence of both $m_{3/2}$ and $M_{\text{SUSY}}$ on the lightest soft term mass $m_{\text{soft}}$, which in turn depends on the mediation mechanism by which SUSY breaking is transmitted to the visible sector. In this work, we focus on gravity and gauge mediation. 
Assuming $m_{\text{soft}}\geq 1\,\text{TeV}$, we obtain\footnote{Throughout this paper we take the 
four-dimensional Planck mass to be $M_{\text{P}} = 2.48\times10^{18}\,\text{GeV}$.}:
\begin{align}
    \label{gravity:bounds}
    \text{gravity mediation:} \qquad 
    & m_{\text{soft}} \sim m_{3/2} \sim \frac{M^2_{\text{SUSY}}}{M_{\text{P}}}, 
    & m_{3/2} \geq 1\,\text{TeV}\,, \\
    \label{gauge:bounds}
    \text{gauge mediation:} \qquad 
    & m_{\text{soft}} \sim \chi\,\frac{m_{3/2} M_{\text{P}}}{M} 
      \sim \chi\,\frac{M^2_{\text{SUSY}}}{M}, 
    & m_{3/2} \geq 0.1\,\text{eV}\,,
\end{align}
where $\chi$ is a coupling of order $\mathcal{O}(10^{-2})$ and $M$ denotes the mediator mass, which we assume to be at least of order  $\mathcal{O}(10)$ TeV. Therefore, the most general lower bound for the gravitino mass, based on current experiments, is 
\begin{equation}
\label{bound:gravitino:mass:LHC}
    m_{3/2}\geq 0.1\,\text{eV}.
\end{equation}
From Eqs.~\eqref{gravity:bounds} and~\eqref{gauge:bounds}, the corresponding bounds on the 
SUSY-breaking scale read:
\begin{align}
    \label{Msusy:gravity}
    \text{gravity mediation:} \qquad & M_{\text{SUSY}} \geq 10^{11}\,\text{GeV}\,, \\[1em] 
    \label{Msusy:gauge}
    \text{gauge mediation:}   \qquad & 10^{4}\,\text{GeV} \leq M_{\text{SUSY}} \leq 10^{11}\,\text{GeV}\,,
\end{align}
where the upper bound for gauge mediation is obtained assuming that the gauge contribution dominates over the gravity contribution by at least a factor of $10^3$. 

\subsection{Relating the two bounds}
\label{chap:gravitino:conj:extra:dimensions}

We now turn to the relation between the gravitino mass and the size of the large extra dimensions. We consider the Gravitino Conjecture and in particular Eq.~\eqref{GravConjFinal}, where, for simplicity, we set the proportionality parameter $\lambda_{3/2}=1$, unless explicitly stated otherwise. The parameter $n $, on the other hand, is the quantity for which we aim to establish constraints.

Using the bounds on the size of extra dimensions\footnote{Throughout the paper we adopt the conversion factor 
$1\,\text{GeV}=5.06\times10^{15}\,\text{m}^{-1}$ between mass and length scales.} reported in table~\ref{tab:correct:astrophisical:bounds:on:l} and in Eq.~\eqref{experimental:bound:on:l}, and applying the GC without assuming any specific dependence of $m_{3/2}$ on the internal volume, we can translate the constraints on the size of the extra dimensions into corresponding bounds on the gravitino mass. In particular, we employ the upper bound from torsion-balance experiments given in Eq.~\eqref{experimental:bound:on:l}, rather than the first entry of table~\ref{tab:correct:astrophisical:bounds:on:l}, since it is more stringent when $p=1$. This procedure yields lower bounds on $m_{3/2}$, which are summarized in table~\ref{tab:bounds:m32:n:p} as functions of $n$ and $p$. 
We consider values of $n$ in the range $0.5 \leq n \leq 2$, consistent with the constraints, which we will derive in Section~\ref{subsec:bounds:n:p}.

\begin{table}[!ht]
    \centering
    \renewcommand{\arraystretch}{1.3}
    \begin{tabular}{|c|c|c|c|c|}
        \hline
        \multicolumn{5}{|c|}{\textbf{Lower bounds on} $m_{3/2}\, [\mathrm{GeV}]$} \\
        \hline
        \textbf{$p$ } & \multicolumn{4}{c|}{\textbf{$n$ }} \\
        \cline{2-5}{}
        & 0.5 & 1 & 1.5 & 2 \\
        \hline
        1 & \cellcolor{gray!45}$1.09\times10^{-41}$ & \cellcolor{gray!45}$5.20\times10^{-12}$ & $4.02\times10^{-2}$ & $3.56\times10^{3}$ \\
        2 & \cellcolor{gray!45}$1.71\times10^{-38}$ & $2.06\times10^{-10}$ & $4.72\times10^{-1}$ & $2.26\times10^{4}$ \\
        3 & \cellcolor{gray!45}$1.21\times10^{-32}$ & $1.73\times10^{-7}$ & $4.21\times10^{1}$ & $6.56\times10^{5}$ \\
        4 & \cellcolor{gray!45}$1.08\times10^{-29}$ & $5.17\times10^{-6}$ & $4.05\times10^{2}$ & $3.58\times10^{6}$ \\
        5 & \cellcolor{gray!45}$6.70\times10^{-28}$ & $4.07\times10^{-5}$ & $1.60\times10^{3}$ & $1.01\times10^{7}$ \\
        6 & \cellcolor{gray!45}$1.08\times10^{-26}$ & $1.63\times10^{-4}$ & $4.04\times10^{3}$ & $2.01\times10^{7}$ \\
        \hline
    \end{tabular}
    \caption{Lower bounds for $m_{3/2}$ in GeV for different values of $p$ and $n$, derived from KK tower scales consistent with current experimental limits. Dark gray cells indicate cases where astrophysical bounds are weaker than the LHC limit, $m_{3/2} \geq 0.1~\text{eV}$. This occurs for $n=0.5$ for all $p$, and for $n=1$, $p=1$.
}
    \label{tab:bounds:m32:n:p}
\end{table}
By combining the results of table~\ref{tab:bounds:m32:n:p} with the experimental lower bound on $m_{3/2}$ from the non-observation of SUSY at the LHC \eqref{bound:gravitino:mass:LHC}, namely $m_{3/2} \geq 0.1\,\text{eV}$, we find that for $n=0.5$ none of the bounds in table~\ref{tab:bounds:m32:n:p} is relevant. A similar situation occurs for $n=1$ with $p=1$. For later convenience, it is also helpful to express the gravitino mass directly in terms of the size of the extra dimensions as
\be
    \label{length:gravconj:equality}
    m_{3/2} = \lambda_{3/2}^{-\frac{1}{n}}\left(\frac{\textrm{m}}{R}\right)^{\frac{1}{n}} \left(\frac{(2.48)^{n-1}}{5.06}\right)^{\frac{1}{n}}10^{\frac{18n-33}{n}}\,\textrm{GeV}\,,
\ee 
where $\textrm{m}$ are meters and $R=m_{\text{KK}}^{-1}$.

The lower bounds obtained from table~\ref{tab:bounds:m32:n:p} represent the strongest constraints on the parameter space of $m_{3/2}$ that can be derived from current upper limits on the size of large extra dimensions, assuming no internal isometries. Although precise predictions are not possible at this stage, in the following sections we will nevertheless establish both upper and lower bounds on $m_{3/2}$ (and on the size of the large extra dimensions).
These constraints will be further sharpened by exploiting the relation between the gravitino mass and the volume of the large $p$-cycle in the compactification manifold.

\section{Decompactifying anisotropic large extra dimensions}\label{section:compactification:generalities}

In this section, we study four-dimensional \(\mathcal{N}=1\) supergravity theories
arising from Type IIA and Type IIB compactifications on Calabi--Yau orientifolds.\footnote{
Our conventions for Type IIA and Type IIB Calabi--Yau orientifold compactifications
are summarized in Appendix~\ref{AppII}; see also
\cite{grimm_effective_2004,louis_generalized_2005,grana_string_2017} for the
corresponding \(\mathcal{N}=1\) effective actions. Appendix~\ref{sec:m_kk:compactification}
contains our conventions for Kaluza--Klein scales, Planck scales, and the effective
KK tower in anisotropic compactifications.} The ten-dimensional spacetime factorizes as
\begin{equation}
\label{product:space:time}
    \mathcal{M}_4 \times Y_6 \,,
\end{equation}
where \(\mathcal{M}_4\) is a non-compact four-dimensional Minkowski spacetime preserving
Poincar\'e invariance, and \(Y_6\) is a Calabi--Yau threefold endowed with an orientifold
involution.

We focus on the case in which the volume of a \(p\)-cycle become large in the internal Calabi--Yau manifold. The associated $p$ directions are related parametrically to the remaining \(6-p\) internal directions through the anisotropy parameter \(\alpha\), defined
asymptotically by the scaling of the total volume with respect to the $p$-cycle one: \(\mathcal V\sim \mathcal V_p^\alpha\). This parametrization affects the effective KK tower and the species
scale, as reviewed in Appendix~\ref{sec:m_kk:compactification}.

The goal of this section is to determine how rewriting the full internal volume in terms
of \(\mathcal V_p\) and $\alpha$ modifies the volume-dependent part of the Type IIB and Type IIA SUGRA K\"ahler potentials. We will show that the relevant logarithmic prefactors
become
\begin{equation}
    K^{\mathcal V}_{IIA}
    =
    -\alpha\ln\mathcal V_p,
    \qquad
    K^{\mathcal V}_{IIB}
    =
    -2\alpha\ln\mathcal V_p .
\end{equation}
These prefactors are the basic input for the subsequent analysis.

Throughout this section we remain agnostic about the detailed geometric realization and
stabilization of the \(p\)-cycle, as well as about the microscopic origin of supersymmetry
breaking. Our analysis is asymptotic and formulated at the level of the four-dimensional
effective field theory. We assume that an F-term is generated and lifts the no-scale vacuum,
giving rise to a finite gravitino mass. We do not address whether the full scalar
potential is compatible with the observed value of the cosmological constant; for related
discussions see, for example, \cite{Hebecker:2019csg}. We also neglect possible warping
effects induced by the anisotropy\footnote{See \cite{Blumenhagen_2020, Blumenhagen_2023} for applications of this in the context of the Swampland program.}.

\subsection{Kähler potentials from $p$-cycle volume} \label{sec:eff:theory:and:meso:extra:dim}

In this section we express the volume-dependent part of the K\"ahler potential in terms of
the large \(p\)-cycle volume \(\mathcal V_p\). The remaining \(6-p\) directions need not
scale trivially along the decompactification trajectory, and their contribution to the full
internal volume is parametrized by the anisotropy exponent \(\alpha\), through
\(\mathcal V\sim \mathcal V_p^\alpha\). We then derive the corresponding logarithmic prefactors for Type IIA
and Type IIB compactifications and show that both exhibit a characteristic scaling.

For Type IIB string theory compactified on a Calabi-Yau orientifold, the resulting four-dimensional effective action is described by $\mathcal{N}=1$ supergravity. The Kähler potential of this effective theory can be written as \cite{grimm_effective_2004}
\begin{equation}
\label{r}
    K_B =  \underbrace{-\ln \left[-i\int_Y \Omega(z) \wedge \bar{\Omega}(\bar{z})\right]}_{K^{cs}_B(z,\bar{z})}\underbrace{-\ln \left[-i(S -\bar{S})\right]-2\ln \left[\frac{1}{6}\mathcal{K}(S,T,G)\right]}_{K^Q_B(S,T,G)}.
\end{equation}

The first term, $K_B^{\rm cs}$, governs the complex structure moduli $z^i$ of $Y$, through the holomorphic three-form $\Omega(z)$. The remaining terms, $K_B^Q$, describe the axio-dilaton $S = C_0 + i e^{-\phi}$ and the Kähler moduli $T$ (and possibly additional moduli $G$ arising from D-branes or fluxes). The Calabi-Yau volume is $\textrm{Vol}(Y) = \frac{1}{6} \mathcal{K}$, with $\mathcal{K} = \mathcal{K}_{\alpha \beta \gamma} v^\alpha v^\beta v^\gamma$ the triple intersection numbers of the two-cycles $v^\alpha$. These terms together define the Kähler potential for the four-dimensional effective theory in Type IIB compactification.

For Type IIA string theory compactified on a Calabi-Yau orientifold, the four-dimensional effective action is again constrained by $\mathcal{N}=1$ supergravity. The Kähler potential can be written as \cite{Grimm:2005aa}
\begin{equation}
\label{Rr}
    K_A = \underbrace{-2\ln \left[2\int_Y \textrm{Re}(C\Omega) \wedge \ast \textrm{Re}(C\Omega) \right]}_{K^Q_A} \underbrace{-\ln \left[\frac{4}{3} \int_Y J \wedge J \wedge J\right]}_{K^k_A}
\end{equation}
where $\int_Y J \wedge J \wedge J = \mathcal{K}_{\alpha \beta \gamma} v^\alpha v^\beta v^\gamma$ gives the Calabi-Yau volume in terms of the two-cycle volumes $v^\alpha$. The first term depends on the four-dimensional \textit{dilaton} through the rescaled holomorphic three-form $C\Omega$, with $C$ a compensator that ensures the correct normalization, while the second term depends on the \textit{Kähler moduli} via the Calabi-Yau volume. For practical purposes, one can thus separate the potential into a \textit{volume-dependent part} ($K^k_A$) and a \textit{dilaton-dependent part} ($K^Q_A$), in analogy with the Type IIB case where the corresponding contributions are $-2 \ln \big[\frac{1}{6}\mathcal{K}\big]$ and $-\ln[-i(S-\bar{S})]$ inside $K_B^Q$.

More details are given in Appendix~\ref{AppII}.

\subsection*{Volume sector and anisotropy dependence}
\label{Sec:anisotropy}
We now introduce a parametrization of the internal anisotropy induced by the presence of
$p$ large directions. Since the scaling of the remaining $6-p$ internal directions is not
fixed a priori, we encode this information in a parameter $\alpha$, allowing us to express
the full internal volume in terms of the large $p$-cycle volume $\mathcal V_p$. Let
$\mathcal V$ denote the full six-dimensional internal volume in string units, and let
$\mathcal V_p$ denote the volume in string units of the $p$-dimensional cycle associated
with the large directions. We define the anisotropy parameter $\alpha$ by
\begin{equation}
\label{Scaling:Alpha}
    \mathcal V \sim \mathcal V_p^\alpha .
\end{equation}
This relation should be understood as an asymptotic scaling relation along the
decompactification trajectory. The parameter $\alpha$ measures how strongly the full
six-dimensional volume grows when the $p$-cycle volume is taken large.

To determine the allowed range of $\alpha$, assume that the $p$ large directions have
a common radius $R$, while the remaining $6-p$ directions have a common radius $r$.
In string units,
\begin{equation}
\label{Vp:def}
    \mathcal V_p =
    \left(
    \frac{R}{\ell_s}
    \right)^p ,
\end{equation}
whereas
\begin{equation}
\label{Vperp:def}
    \mathcal V_{6-p} =
    \left(
    \frac{r}{\ell_s}
    \right)^{6-p}.
\end{equation}
The full internal volume is therefore
\begin{equation}
\label{volume:decomposition}
    \mathcal V
    =
    \mathcal V_p\,\mathcal V_{6-p}
    =
    \left(
    \frac{R}{\ell_s}
    \right)^p
    \left(
    \frac{r}{\ell_s}
    \right)^{6-p}.
\end{equation}
Combining \eqref{Scaling:Alpha} with \eqref{Vp:def} and
\eqref{volume:decomposition}, one obtains
\begin{equation}
\label{rsrl}
    \frac{r}{\ell_s}
    =
    \left(
    \frac{R}{\ell_s}
    \right)^{\frac{p(\alpha-1)}{6-p}} .
\end{equation}
We require the transverse directions to be no smaller than the string length and no larger
than the large directions,
\begin{equation}
\label{r:range}
    \ell_s \leq r \leq R .
\end{equation}
For $R\gg \ell_s$, this condition implies
\begin{equation}
\label{Alpha:range}
    1 \leq \alpha \leq \frac{6}{p}.
\end{equation}

This immediately implies
\begin{equation}
\label{range:alpha:p}
    p \leq \alpha p \leq 6\,,
\end{equation}
where the combination $\alpha p$ measures the effective number of internal directions
participating in the decompactification limit. The two endpoints have a simple geometric interpretation:
\begin{itemize}
    \item For $\alpha=1$, one finds $r\simeq \ell_s$. This is the maximally
    anisotropic regime: the $p$ directions inside $\mathcal V_p$ become large, while
    the remaining $6-p$ directions remain at string size. In this case, we have that $\alpha p=p$ and
    \begin{equation}
        \mathcal V \sim \mathcal V_p\,.
    \end{equation}

    \item For $\alpha=6/p$, one finds $r=R$. This is the isotropic endpoint, in
    which all six internal directions scale with the same radius. In this case, $\alpha p=6$ and
    \begin{equation}
        \mathcal V \sim \mathcal V_p^{6/p}\,.
    \end{equation}
\end{itemize}

We can now determine how this anisotropy affects the volume-dependent part of the
K\"ahler potential. Let us first recall how the volume appears in the K\"ahler sector.

In Type IIB compactifications, the internal volume depends implicitly on the real moduli
$v_k$ through the real part of the K\"ahler coordinates $T_k$, denoted $\tau_k$. Here,
the index $k=1,\dots,h^{1,1}$ runs over the independent four-cycles of the Calabi--Yau,
with each $\tau_k$ corresponding to the volume of a distinct four-cycle. The $\tau_k$
and $v_k$ are related as in equation \eqref{relationFourcylesTwocycles}. The full
Calabi--Yau volume is a homogeneous function of degree $3/2$ in the $\tau_k$,
\begin{equation}
\label{GenVolumeModuliIIB}
    \mathcal V \sim f_{3/2}(\tau_k) .
\end{equation}
In the single-modulus scaling approximation, this reduces to
\begin{equation}
    \mathcal V \sim \tau^{3/2}.
\end{equation}
Combining this relation with \eqref{Scaling:Alpha}, one obtains\footnote{
In Type IIB the chiral K\"ahler coordinates are naturally associated with
four-cycle volumes, while the K\"ahler form is expanded in two-cycle moduli.
Therefore, a literal cycle interpretation of $\mathcal V_p$ is most direct for
even-dimensional cycles. In this work, however, $\mathcal V_p$ denotes the
effective volume spanned by $p$ large real directions and is used as an
asymptotic scaling variable for $p=1,\ldots,6$.
}
\begin{equation}
\label{Alpha:tau}
    \tau
    \sim
    \mathcal V^{2/3}
    \sim
    \mathcal V_p^{2\alpha/3}.
\end{equation}

In Type IIA compactifications, instead, the volume is naturally expressed in terms of
the two-cycle moduli $v_i$ appearing in the expansion of the K\"ahler form. Here the index
$i=1,\dots,h^{1,1}$ runs over the independent two-cycles. As reviewed in
Appendix~\ref{AppII}, the full internal volume is a homogeneous cubic function of these
variables,
\begin{equation}
\label{GenVolumeModuliIIA}
    \mathcal V \sim f_3(v_i).
\end{equation}
In the corresponding single-modulus scaling approximation,
\begin{equation}
    \mathcal V \sim v^3 .
\end{equation}
Using again \eqref{Scaling:Alpha}, we find
\begin{equation}
\label{Alpha:v}
    v
    \sim
    \mathcal V^{1/3}
    \sim
    \mathcal V_p^{\alpha/3}.
\end{equation}

Our main interest is in the logarithmic prefactors controlling the scaling with
$\mathcal V_p$, since these are the quantities modified by the anisotropy parameter
$\alpha$. For a single K\"ahler modulus, the relevant logarithmic dependence can be written
as\footnote{This is the most general K\"ahler potential producing the metric of a space with
Poincar\'e invariance. The prefactor $3$ is fixed by requiring canonical kinetic terms
\cite{freedman_supergravity_2013}.}
\begin{equation}
\label{K:single:modulus}
    K=-3\ln(\rho+\bar\rho),
\end{equation}
where $\rho+\bar\rho$ is identified with the real modulus controlling the internal
volume. In Type IIB, this real modulus is the four-cycle volume $\tau$. Therefore,
using \eqref{Alpha:tau},
\begin{equation}
\label{K:B:vol}
    K^{\mathcal V}_{B}
    =
    -3\ln\tau
    =
    -3\ln\!\left(\mathcal V_p^{2\alpha/3}\right)
    =
    -2\alpha\ln\mathcal V_p .
\end{equation}
In Type IIA, the corresponding real modulus is the two-cycle volume $v$. Using
\eqref{Alpha:v}, one obtains
\begin{equation}
\label{K:A:vol}
    K^{\mathcal V}_{A}
    =
    -3\ln v
    =
    -3\ln\!\left(\mathcal V_p^{\alpha/3}\right)
    =
    -\alpha\ln\mathcal V_p .
\end{equation}

Thus the volume-dependent contributions to the K\"ahler potentials are
\begin{equation}
\label{Alpha:K:vol}
\boxed{
    K^{\mathcal V}_{B}
    =
    -2\alpha\ln\mathcal V_p\,,
    \qquad
    K^{\mathcal V}_{A}
    =
    -\alpha\ln\mathcal V_p\,,
}
\end{equation}
respectively for Type IIB and Type IIA compactifications, up to additive constants. These are the expressions that will be used in the following sections. They show that the
dependence of the four-dimensional effective theory on the $p$-cycle volume is controlled
not only by the number $p$ of large directions, but also by the anisotropy parameter
$\alpha$. Equivalently, the usual dependence on the full internal volume is translated into
a dependence on $\mathcal V_p$, with $\alpha$ encoding how the remaining internal
directions scale along the decompactification trajectory.

Several consistency checks follow immediately. At the maximally anisotropic endpoint
$\alpha=1$, one has $\mathcal V\sim\mathcal V_p$, and therefore
\eqref{Alpha:K:vol} reduces to the standard large-volume expressions
\begin{equation}
\label{K:vol:standard}
    K^{\mathcal V}_{B}
    =
    -2\ln \mathcal V\,,
    \qquad
    K^{\mathcal V}_{A}
    =
    -\ln \mathcal V\,.
\end{equation}
At the isotropic endpoint $\alpha=6/p$, instead,
\begin{equation}
    K^{\mathcal V}_{B}
    =
    -\frac{12}{p}\ln\mathcal V_p\,,
    \qquad
    K^{\mathcal V}_{A}
    =
    -\frac{6}{p}\ln\mathcal V_p\,.
\end{equation}
These are precisely the prefactors obtained by rewriting the full volume in terms of
$\mathcal V_p$ under the isotropic scaling $\mathcal V_p\sim\mathcal V^{p/6}$. Hence
the $p$-dependent prefactors arise as a special isotropic limit, while the general
anisotropic dependence is controlled by $\alpha$.

\subsection*{Dilaton sector and mirror relations}\label{subsect:dilaton}

The dilaton-dependent sector of the Kähler potential requires a separate discussion
with respect to the volume sector.
Unlike the latter, it is not directly associated with the expansion of an internal
$p$-cycle, which we have assumed to be governed exclusively by Kähler moduli.
In particular, in Type IIB compactifications the relevant contribution $K_B^{Q}$
depends on the axio-dilaton $S$, while in Type IIA it is encoded in the
complex-structure sector.

For Type IIB, one finds \cite{Grimm:2005aa}
\begin{equation}
\label{KQB:IIB}
K^Q_{B}(S,G,T)
= -2 \ln \!\left[e^{-2 \phi^B_{(10)}} \int J \wedge J \wedge J\right]
= -\ln\!\left(e^{-4\phi^B_{(4)}}\right)\,,
\end{equation}
where $\phi^B_{(10)}$ and $\phi^B_{(4)}$ denote the ten- and four-dimensional dilatons,
respectively.
The corresponding quantities in Type IIA are obtained by replacing the superscript
$B$ with $A$.
At the level of the four-dimensional effective theory, approximate
$\mathcal{N}=1$ mirror symmetry implies the universal relation
\begin{equation}
\label{mirror:relation}
K^Q_B = K^Q_A = 4\phi_{(4)} \,,
\qquad
\phi_{(4)}=\phi^B_{(4)}=\phi^A_{(4)}\, .
\end{equation}
This identification, however, relies on a specific choice of frame for the Kähler moduli.
In Type IIB, the expression \eqref{KQB:IIB} is naturally obtained when the real moduli $v$ are written in the string frame, related to their Einstein-frame counterparts by $v^{(E)} \rightarrow e^{-\frac12\phi_B} v^{(S)}$. This rescaling is independent of the number of large internal dimensions.

Making this conversion explicit and using $\mathrm{Im}\,S \sim g_s^{-1}$,
one finds
\begin{align}
\label{KQB:general}
K^Q_B
&= -\ln\!\left(g_s^{-1}\right)
   - 2\alpha \ln\!\left(\mathcal{V}^{(E)}_{p}\right)
   = -4\ln\!\left(g_s^{-1}\right)
   - 2\alpha \ln\!\left(\mathcal{V}^{(S)}_{p}\right)\,,
\end{align}
where the additional dilaton dependence arises precisely from the rescaling of the Kähler moduli to string frame.

It is instructive to rewrite the Type IIB result in terms of the four-dimensional string
coupling. Using the relation between the ten- and four-dimensional string couplings $g_s$ and $g_{(4)}$ 
(see Appendix~\ref{sec:m_kk:compactification}, eq.~\eqref{dilatons}), one can equivalently express
\begin{align}
K^Q_B
&= -4\ln\!\left(g^{-1}_{(4)}\right)
   - \left(2\alpha -2\alpha \right)\ln\!\left(\mathcal{V}^{(S)}_{p}\right).
\end{align}

The second term vanishes for every $\alpha$, hence the standard mirror-symmetric relation $K^Q_B = K^Q_A = 4\phi_{(4)}$ is exactly recovered.

\section{Refining the Gravitino Conjecture} \label{sec:volume:relations}

In this section, we refine the Gravitino Conjecture by making explicit the
dependence of its scaling exponent $n$ on the anisotropy parameter $\alpha$ and the number $p$ of large extra dimensions. 
To this end, we determine how both the gravitino mass $m_{3/2}$ and the
Kaluza--Klein scale $m_{\text{KK}}$ scale with the volume $\mathcal{V}_p$. We work throughout in perturbative string theory and carefully distinguish
between Type IIA and Type IIB compactifications, for which the appropriate
perturbativity conditions are naturally expressed in terms of the
four-dimensional coupling $g_{(4)}$ or the ten-dimensional coupling $g_s$,
respectively, following \cite{Font_2019}.

The section is organized as follows. We first review the conditions for
perturbative control in the large-volume limit. We then derive the anisotropy-dependent volume scaling of the Kaluza--Klein scale and of the gravitino mass. Finally, we impose the Gravitino Conjecture together with consistency constraints from the species scale to obtain $\alpha p$-dependent bounds on the exponent $n$.

\subsection{Perturbativity and large volume limit}
\label{sec:perturbative:regime:string:theory}

To consistently compare Type IIA and Type IIB compactifications, it is important to express the relevant quantities in terms of the appropriate string coupling: for Type IIA, in terms of the four-dimensional dilaton $g_{(4)}$, and for Type IIB, in terms of the ten-dimensional dilaton $g_s = e^{\phi_{(10)}}$. 
As emphasized in \cite{Font_2019}, this choice identifies the regions of moduli space where a controlled effective field theory exists, particularly in the large-volume limit, and imposes the universal perturbativity condition
\begin{equation}
    g_s < 1\,.
\end{equation}
Here, $g_s$ is not treated as a continuously vanishing parameter, but rather as a small fixed quantity, with decompactification limits considered only in the asymptotic regime $m_{3/2} \rightarrow 0$.

In Type IIA, the internal volume $\mathcal{V}_p$ enters the Kähler potential through the standard volume-dependent term $K^K$, while the quaternionic sector encodes the four-dimensional dilaton via $K^Q = 4\phi_{(4)}$, as it was already outlined with Eq.~\eqref{Rr}. The ten-dimensional dilaton is then related to the four-dimensional one as $e^{\phi_{(10)}} = e^{\phi_{(4)}} \sqrt{\mathcal{V}^\alpha_p}$, or equivalently
\begin{equation}
\label{IIA:dilaton}
e^{\phi_{(10)}} \propto e^{K^Q/4} \, e^{-K^K/2} \,,
\end{equation}
where the second factor arises from the Kähler moduli contribution. In the large-volume limit, one or more Kähler moduli diverge, and staying in the perturbative regime requires that $e^{K^Q/4}$ scales appropriately with $e^{-K^K/2}$, leading to
\begin{equation}
\label{Pert:IIA}
    e^{K^Q/4} \sim g_{(4)} < 1\,.
\end{equation}

By contrast, in Type IIB the internal volume enters the Kähler moduli part of the quaternionic Kähler potential. The four-dimensional dilaton $\phi_{(4)}$ also originates from the quaternionic sector, in analogy with Type IIA under mirror symmetry. In the Type IIB case, however, this sector naturally splits into an axio-dilaton contribution $K^S$ and a Kähler moduli contribution $K^K$, so that $K^Q = K^S + K^K$. The axio-dilaton chiral multiplet $S$ is defined in ten dimensions as
$S = C_0 + i\, e^{-\phi_{(10)}}$, and therefore $K^S$ captures the full dependence on the ten-dimensional dilaton. The ten-dimensional string coupling $g_s = e^{\phi_{(10)}}$ can be expressed in terms of the four-dimensional Kähler potential as
\begin{equation}
\label{IIB:dilaton}
e^{\phi_{(10)}} \propto e^{K^S/4} \, e^{-K^K/4} \,,
\end{equation}
where the first factor originates from the axio-dilaton sector, while the second reflects the rescaling induced by the Kähler moduli contribution. Remaining in the perturbative regime then requires
\begin{equation}
    e^{K^S/4} \sim g_s < 1\,,
\end{equation}
independently of the size of the internal volume.

Mirror symmetry provides a nontrivial duality between Type IIA and Type IIB compactifications, mapping the large-volume limit of IIA to the large-complex-structure limit of IIB. In this duality, the four-dimensional dilaton arises differently on the two sides, which explains why different frame choices or moduli normalizations are necessary to consistently match physical quantities across the mirror pair. In particular, this justifies why the perturbative string coupling is identified with $g_{(4)}$ in IIA and with $g_s$ in IIB, and how these choices ensure a controlled EFT in the large-volume regime.

Finally, as additional support for assuming asymptotic decompactification within perturbative string theory, consider that equation \eqref{Pert:IIA} enforces an inverse scaling between the four-dimensional dilaton $g_{(4)}$ and the internal volume. In these limits, the required co-scaling between the dilaton and the internal volume is violated, resulting in a decompactification regime that does not admit a controlled interpretation from the M-theory perspective \cite{Lee:2019wij}.

Since our focus is on the large volume regime, we move in asymptotic regions of the $\mathcal{N}=1$ moduli space of a Type II Calabi-Yau orientifold compactifications, viewed as a consistent truncation of the underlying $\mathcal{N}=2$ moduli space. We investigate these limits in the presence of spontaneously broken supersymmetry, relying on the fact that turning on a scalar potential does not change the asymptotic exponential rates of the towers and strings \cite{Grieco:2025bjy}. However, it may obstruct certain infinite-distance limits via divergences or quantum corrections \cite{Grieco:2025bjy}, but the precise study of this issue is beyond the aim of this work.

\subsection{Kaluza--Klein mass and $p$-cycle volume dependence}

In the two-tower example considered, the ratio between the $p$-cycle and the $(6-p)$-cycle in the internal volume is fixed by the volume decomposition \eqref{volume:decomposition} and by \eqref{Scaling:Alpha}. The lightest Kaluza--Klein scale in a regime where only $p$ extra dimensions are large is set by the string scale $M_s$ and the corresponding $p$-cycle volume as
\begin{equation}
\label{mkk:def}
    m_{\text{KK}} = \frac{M_s}{\mathcal V_p^{1/p}} \,.
\end{equation}
The effective tower is instead given by \eqref{effective:tower:example} as 
\be
\label{eff:mkk}
m_\text{KK,eff}= \frac{M_s}{\mathcal{V}_p^{\alpha/6}}\,.  
\ee
These expressions show the explicit geometric dependence on $\mathcal V_p$ and $\alpha$. Any additional volume or dilaton dependence of the KK scale enters only through the string scale $M_s$.

Working in four-dimensional Einstein frame, the string scale is related to the four-dimensional Planck mass $M_{\text{P}}$ through the dilaton and the internal volume. Schematically, this relation takes the form
\begin{equation}
\label{string mass:IIB}
    M_s^2 \sim g_{(4)}^2 \, M_{\text{P}}^2 \sim \frac{g_s^2}{\mathcal V^\alpha_p} \, M_{\text{P}}^2 \,,
\end{equation}
where we have first expressed $M_s$ in terms of the four-dimensional dilaton and subsequently rewritten it in terms of the ten-dimensional string coupling $g_s = e^{\phi_{(10)}}$. Numerical factors of $2\pi$ have been omitted throughout and can be consistently absorbed into the constant $\lambda_{3/2}$, set to unity in the present derivation.

Combining Eqs.~\eqref{mkk:def} and \eqref{eff:mkk} with \eqref{string mass:IIB}, and expressing all mass scales in four-dimensional Planck units, we can write the KK scales entirely in terms of the cycle volume $\mathcal V_p$ and the perturbative string coupling appropriate to the compactification under consideration. As discussed in Section~\ref{sec:perturbative:regime:string:theory}, perturbative control is achieved in terms of the four-dimensional coupling $g_{(4)}$ in Type IIA and the ten-dimensional coupling $g_s$ in Type IIB. This leads to 
\begin{equation}
\label{Mkk:volume:full:dilaton:relations}
\begin{aligned}
    \text{IIA:} \quad & m_{\text{KK}} \sim g_{(4)} \, \mathcal V_p^{-\frac{1}{p}} \,, && \quad && m_{\text{KK},\text{eff}} \sim g_{(4)} \, \mathcal V_p^{-\frac{\alpha}{6}}\,,\\
    \text{IIB:} \quad & m_{\text{KK}} \sim g_s \, \mathcal V_p^{-\left(\frac{\alpha}{2} + \frac{1}{p}\right)} \,, && \quad && m_{\text{KK},\text{eff}} \sim g_s \, \mathcal V_p^{-\left(\frac{\alpha}{2} + \frac{\alpha}{6}\right)} \,.
\end{aligned}
\end{equation}

The additional $\mathcal V_p^{-\alpha/2}$ dependence appearing in the Type IIB case originates from the relation between the four- and ten-dimensional dilatons under dimensional reduction. In Einstein frame, expressing the string scale in terms of the ten-dimensional coupling $g_s$ introduces an extra volume factor relative to the case in which the effective theory is written in terms of the four-dimensional coupling $g_{(4)}$, as in Type IIA.

In the remainder of this section, all mass scales will be expressed in four-dimensional Planck units.

\subsection{Gravitino mass and $p$-cycle volume dependence}
\label{subsubsect:GravitinoMassAndVolumeDependence}

In this subsection we determine how the gravitino mass $m_{3/2}$ scales with the
volume of the internal compactification manifold in the presence of $p$ large
extra dimensions. Our goal is to make explicit the dependence on $\mathcal{V}_p$
and on the string coupling, distinguishing between Type IIA and Type IIB
compactifications, and to prepare the ground for a comparison with the Kaluza--Klein
scale relevant for the Gravitino Conjecture.

We recall from the standard $\mathcal{N}=1$ supergravity expression for the gravitino
mass,
\begin{equation}
m_{3/2} = e^{K/2} |W| \,,
\label{field:dependent::gravitino:mass2}
\end{equation}
and focus only on the volume- and dilaton-dependent contributions to the Kähler
potential arising in Type II compactifications, as discussed in
Section~\ref{Sec:anisotropy}.

Before proceeding, we must specify how the superpotential scales with the internal
volume. Since we are interested in the Kähler-moduli dependence and assume that all
other moduli are stabilized, we parametrize the vacuum expectation value of the
superpotential as
\begin{equation}
\label{volume:superpotential}
\langle W \rangle \sim \mathcal{V}_p^{\beta/2} \,,
\end{equation}
where $\beta$ is a model-dependent parameter encoding the asymptotic volume scaling.
Possible dilaton-dependent contributions can be included by introducing an
additional exponent $k$, which parametrizes $g_s$ terms in Type IIB and $g_{(4)}$ in Type IIA, as
\be
\bal
\label{superpotential:vol:and:dil}
\textrm{IIA:} \quad \langle W \rangle \sim \mathcal{V}_p^{\beta/2} g_{(4)}^{k}\,, && \qquad &&\textrm{IIB:} \quad\langle W \rangle \sim \mathcal{V}_p^{\beta/2} g_s^{k}\,.
\eal
\ee

Combining the volume-dependent parts of the Kähler potentials for Type IIA and Type
IIB with Eq.~\eqref{superpotential:vol:and:dil} and using
Eq.~\eqref{field:dependent::gravitino:mass2}, we obtain the leading volume and
dilaton dependence of the gravitino mass. Expressing the result in four-dimensional
Planck units and imposing the appropriate perturbative regime discussed in
Section~\ref{sec:perturbative:regime:string:theory}, we find
\begin{equation}
\label{volume:m3/2:full:dilaton}
\begin{aligned}
\textrm{IIA:} \qquad
m_{3/2} &\sim \mathcal{V}_p^{-\frac{\alpha}{2}+\frac{\beta}{2}} \, g_{(4)}^{2+k} \,, \\
\textrm{IIB:} \qquad
m_{3/2} &\sim \mathcal{V}_p^{-\alpha+\frac{\beta}{2}} \, g_s^{\frac{1}{2}+k} \,.
\end{aligned}
\end{equation}

Let us now focus on the volume dependence and temporarily disregard dilaton-dependent factors. In Type IIB flux compactifications, a standard tree-level assumption is that the superpotential is independent of the Kähler moduli, corresponding to $\beta=0$. In this case, Kähler-moduli dependence arises only through non-perturbative effects. In Type IIA flux compactifications, by contrast, Ramond--Ramond fluxes generically
induce a tree-level dependence of $W$ on the Kähler moduli, suggesting $\beta \neq 0$ as the generic situation. Nevertheless, in the following we will set $\beta=0$ also in the Type IIA case.
This choice allows for a direct comparison with the results of Ref.~\cite{Castellano:2021aa}, where general Calabi--Yau threefold compactifications were analyzed using asymptotic Hodge-theoretic methods. In that work, a lower bound on the exponent $n$ appearing in the Gravitino Conjecture was obtained by comparing
the gravitino mass with the Kaluza--Klein scale in Planck units,
\begin{equation}
m_{3/2}^{\,n_{\text{crit}}} \geq m_{\text{KK}} \,.
\end{equation}

In our framework, the same scaling relation emerges naturally once the explicit
volume dependence of $m_{3/2}$ is taken into account. For a single Kähler modulus
entering the superpotential as $W \sim v^{\gamma}$, one has
$W \sim \mathcal{V}_p^{2\gamma/p}$ and therefore $\beta = 4\gamma/p$.
Setting $\gamma=0$ yields $\beta=0$ and leads to the same lower bound
$n \geq 1/3$ found in Ref.~\cite{Castellano:2021aa}.
This agreement provides a posteriori justification for adopting $\beta=0$ in both
Type IIA and Type IIB compactifications in our analysis.

\subsection{Bounds on the scaling exponent $n$ and $p$-dependence}
\label{subsec:bounds:n:p}

We are now ready to impose the Gravitino Conjecture using the results derived above
for the dependence of both the Kaluza--Klein scale $m_{\text{KK}}$ and the gravitino
mass $m_{3/2}$ on the volume $\mathcal{V}_p$ and on the string coupling.
Combining the relations obtained in the previous subsections, we can express the
gravitino mass directly in terms of the KK scale, keeping explicit track of the
number $p$ of large internal dimensions.

Allowing for a possible volume and dilaton dependence of the superpotential,
parametrized by $\beta$ and $k$, one finds that the gravitino mass depends on the lightest KK mass scale and the string coupling as
\begin{equation}
\label{relations:with:beta}
\begin{aligned}
\text{IIA:}\quad
& m_{3/2} \sim m_{\text{KK}}^{\,\frac{p}{2}(\alpha-\beta)}
\, g_{(4)}^{\,2 + k}\,g_{(4)}^{-\frac{p}{2}(\alpha-\beta)}\,, \\[4pt]
\text{IIB:}\quad
& m_{3/2} \sim m_{\text{KK}}^{\,\frac{p\,(2\alpha - \beta)}{2 + \alpha p}}
\, g_{s}^{\,\frac{1}{2} + k}\,g_{s}^{-\frac{p\,(2\alpha - \beta)}{2 + \alpha p}}\,,
\end{aligned}
\end{equation}
where we have used Eq.~\eqref{Mkk:volume:full:dilaton:relations} and Eq.~\eqref{volume:m3/2:full:dilaton}. In the following, we restrict to the minimal and most conservative case
$\beta = 0$, corresponding to a constant volume-dependence in the superpotential at leading order.

With this simplification, Eq.~\eqref{relations:with:beta} reduces to
\begin{equation}
\label{relation:gravitino:mkk}
\begin{aligned}
\text{IIA:}\quad
& m_{3/2} \sim m_{\text{KK}}^{\,\frac{\alpha p}{2}}\, g_{(4)}^{-\frac{\alpha p}{2} +2 + k}\,, \\[4pt]
\text{IIB:}\quad
& m_{3/2} \sim m_{\text{KK}}^{\,\frac{2\alpha p}{\alpha p+2}}
\, g_s^{-\frac{2\alpha p}{\alpha p+2} + \frac12 + k}\, .
\end{aligned}
\end{equation}
In simple 4d $\mathcal{N}=1$ Type IIA and Type IIB compactifications, the superpotential depends at tree-level on the four- or ten-dimensional dilaton, respectively, with a power $k$ which can be either $k=0$ or $k=-1$ \cite{Camara:2005dc}. In particular, assuming $\beta=0$ in Type IIA amounts to neglecting RR-flux contributions to the superpotential, and retaining only NS and geometric fluxes. Crucially, these generically generate superpotential terms at tree-level that scale with the four-dimensional dilaton as $k=-1$\cite{Camara:2005dc}. Therefore, it is natural to consider $k=-1$ for Type IIA, which is equivalent to assume $\beta=0$. 

With these values, the power of the string coupling appearing in
Eq.~\eqref{relation:gravitino:mkk} is always negative for Type IIB. Instead, the condition $k=-1$ in Type IIA does not, in general, ensure a negative overall exponent of the four-dimensional string coupling. In the relation below, we assume that this negativity condition is satisfied.

Imposing perturbative control requires $g_{(4)} < 1$ in Type IIA and
$g_s < 1$ in Type IIB. This seems to suggest the inequalities
\begin{equation}
\begin{aligned}
\text{IIA:}\quad & m_{3/2} \geq m_{\text{KK}}^{\frac{\alpha p}{2}}\,,
\qquad
\text{IIB:}\quad & m_{3/2} \geq m_{\text{KK}}^{\frac{2\alpha p}{\alpha p+2}}\, ,
\end{aligned}
\end{equation}

Comparing these relations with the Gravitino Conjecture,
$m_{3/2}^{\,n} \sim m_{\text{KK}}$, we obtain lower bounds on the exponent $n$,
\begin{equation}
\label{n:lower:bound:compactification}
\text{IIA:}\quad \frac{2}{\alpha p} \leq n\,, 
\qquad
\text{IIB:}\quad \frac{\alpha p+2}{2\alpha p} \leq n\, .
\end{equation}
The qualitative difference between the lower bounds in Type IIA and Type IIB arises from the distinct role of the string coupling. 

It is important to notice that, in Type IIA, the lower bound is meaningful only when $\alpha p \ge2$, since the power of the four-dimensional string coupling in the relation \eqref{relation:gravitino:mkk} is required to remain non-positive. In particular, for $p=1$, this excludes the range $1 \leq \alpha < 2$, so that no consistent lower bound can be defined there.

An upper bound on $n$ follows from the requirement that supersymmetry breaking must occur below the species scale,
\begin{equation}
    M_{\text{SUSY}} \le \Lambda_{\text{sp}}\,,
\end{equation}
ensuring the consistency of the effective field theory \cite{Anchordoqui:2023oqm}. Using the relation
between the SUSY-breaking scale and the gravitino mass, together with the
expression of the species scale in terms of the KK mass \eqref{species:scale:effective}, this condition leads to
\begin{equation}
\label{strong:n:bound}
n \leq \frac{\alpha p+2}{2 \alpha p}\, .
\end{equation}

Combining the lower and upper bounds, the allowed intervals are
\begin{equation}
\label{bounds:n:IIA:IIB}
\begin{aligned}
\text{IIA:}\quad  \frac{2}{\alpha \,p} \le n \leq \frac{\alpha p+2}{2 \alpha p}\,,
\qquad
\text{IIB:}\quad  n = \frac{\alpha p+2}{2 \alpha p}\,.
\end{aligned}
\end{equation}
For Type IIA with $p=1$ and $1\leq \alpha<2$, \textit{i.e.} $\alpha p < 2$, our analysis, through the formula Eq. \eqref{relation:gravitino:mkk} enforced with the perturbativity condition $g_{(4)}<1$, gives only an upper bound for $n$: $n<\frac{2}{\alpha \, p}$. For $\alpha p < 2$ this is always a less stringent bound than \eqref{strong:n:bound}, therefore, all together our analysis predicts
\be
\label{bound:n:TypeIIA:alpha<2}
\text{IIA:}\quad n \leq \frac{\alpha p+2}{2 \alpha p}\,, \quad \text{with }\alpha\,p<2\,.
\ee

This does not mean that the geometry $p=1$, $\alpha=1$ in Type IIA is not possible, rather that our derivation of the Type IIA bound \eqref{n:lower:bound:compactification} does not apply in this case\footnote{Indeed, in section \ref{sec:mirror:symmetry} we found that every value of $\alpha$ respects the mirror symmetry equivalence of four-dimensional dilatons\eqref{mirror:relation}.}.

\medskip

Interestingly, we can see that the only value of $\alpha p$ for which the Type IIA and Type IIB $n$ values coincide is $\alpha p=2$, for which holds $n_{IIA}=n_{IIB}=1$. This means that the possibilities are either $p=1$ and $\alpha=2$ or $p=2$ and $\alpha=1$.

In table \ref{tab:allowed:n:intervals} we give the allowed values for $n$ for Type IIA and Type IIB obtained from the relations \eqref{bounds:n:IIA:IIB} at fixed $p$ and varying $\alpha$. At fixed $\alpha p$, then \eqref{bounds:n:IIA:IIB} collapses to a single value for Type IIB and lower values of $\alpha p$ will increase the lower bound for Type IIA. Recall that the quantity $\alpha p$, denoting the number of extra dimensions effectively becoming large, varies as \eqref{range:alpha:p}. Bounds on $n$ for fixed $\alpha=1$ and $\alpha=2$ are given in table \ref{tab:allowed:n:fixed:alpha}.

\begin{table}[!htbp]
\centering
\renewcommand{\arraystretch}{1.4}

\begin{tabular}{|c|r@{\;}c@{\;}c@{\;}c@{\;}l|r@{\;}c@{\;}c@{\;}c@{\;}l|}
\hline
\multicolumn{11}{|c|}{\textbf{Bounds on \(n\) for allowed \(\alpha\)}} \\
\hline
\textbf{\(p\)}
& \multicolumn{5}{c|}{\textbf{IIA}}
& \multicolumn{5}{c|}{\textbf{IIB}} \\
\hline
1 & \(1/3\) & \(\leq\) & \(n\) & \(\leq\) & \(1\)
  & \(2/3\) & \(\leq\) & \(n\) & \(\leq\) & \(3/2\) \\

2 & \(1/3\) & \(\leq\) & \(n\) & \(\leq\) & \(1\)
  & \(2/3\) & \(\leq\) & \(n\) & \(\leq\) & \(1\) \\

3 & \(1/3\) & \(\leq\) & \(n\) & \(\leq\) & \(5/6\)
  & \(2/3\) & \(\leq\) & \(n\) & \(\leq\) & \(5/6\) \\

4 & \(1/3\) & \(\leq\) & \(n\) & \(\leq\) & \(3/4\)
  & \(2/3\) & \(\leq\) & \(n\) & \(\leq\) & \(3/4\) \\

5 & \(1/3\) & \(\leq\) & \(n\) & \(\leq\) & \(7/10\)
  & \(2/3\) & \(\leq\) & \(n\) & \(\leq\) & \(7/10\) \\

6 & \(1/3\) & \(\leq\) & \(n\) & \(\leq\) & \(2/3\)
  & \multicolumn{5}{c|}{\(n=2/3\)} \\
\hline
\end{tabular}

\caption{Numerical values of the exponent \(n\) in the Gravitino Conjecture for Type IIA and Type IIB compactifications, as a function of the number \(p\) of $p$-cycle directions, after allowing \(\alpha\) to vary in the ranges \eqref{bounds:n:IIA:IIB}. For Type IIA, we reported only bounds for $\alpha p \geq 2$.}
\label{tab:allowed:n:intervals}
\end{table}

\begin{table}[!htbp]
\centering
\renewcommand{\arraystretch}{1.2}

\begin{minipage}[t]{0.48\textwidth}
\centering
\begin{tabular}{|c|r@{\;}c@{\;}c@{\;}c@{\;}l|c|}
\hline
\multicolumn{7}{|c|}{\textbf{Bounds on \(n\) with \(\alpha=1\)}} \\
\hline
\textbf{\(p\)}
& \multicolumn{5}{c|}{\textbf{IIA}}
& \textbf{IIB} \\
\hline
1 & \multicolumn{5}{c|}{\(n\leq 3/2\)}
  & \(n=3/2\) \\

2 & \multicolumn{5}{c|}{\(n=1\)}
  & \(n=1\) \\

3 & \(2/3\) & \(\leq\) & \(n\) & \(\leq\) & \(5/6\)
  & \(n=5/6\) \\

4 & \(1/2\) & \(\leq\) & \(n\) & \(\leq\) & \(3/4\)
  & \(n=3/4\) \\

5 & \(2/5\) & \(\leq\) & \(n\) & \(\leq\) & \(7/10\)
  & \(n=7/10\) \\

6 & \(1/3\) & \(\leq\) & \(n\) & \(\leq\) & \(2/3\)
  & \(n=2/3\) \\
\hline
\end{tabular}
\end{minipage}
\hfill
\begin{minipage}[t]{0.48\textwidth}
\centering
\begin{tabular}{|c|r@{\;}c@{\;}c@{\;}c@{\;}l|c|}
\hline
\multicolumn{7}{|c|}{\textbf{Bounds on \(n\) with \(\alpha=2\)}} \\
\hline
\textbf{\(p\)}
& \multicolumn{5}{c|}{\textbf{IIA}}
& \textbf{IIB} \\
\hline
1 & \multicolumn{5}{c|}{\(n=1\)}
  & \(n=1\) \\

2 & \(1/2\) & \(\leq\) & \(n\) & \(\leq\) & \(3/4\)
  & \(n=3/4\) \\

3 & \(1/3\) & \(\leq\) & \(n\) & \(\leq\) & \(2/3\)
  & \(n=2/3\) \\

4 & \(1/4\) & \(\leq\) & \(n\) & \(\leq\) & \(5/8\)
  & \(n=5/8\) \\

5 & \(1/5\) & \(\leq\) & \(n\) & \(\leq\) & \(3/5\)
  & \(n=3/5\) \\

6 & \(1/6\) & \(\leq\) & \(n\) & \(\leq\) & \(7/12\)
  & \(n=7/12\) \\
\hline
\end{tabular}
\end{minipage}

\caption{Allowed values of \(n\) for fixed representative values of \(\alpha\).
In the left table, for Type IIA and \(\alpha=1\), the bound is the one in
\eqref{bound:n:TypeIIA:alpha<2}.}
\label{tab:allowed:n:fixed:alpha}
\end{table}

The tables above should be read with an important distinction in mind. At fixed
\(\alpha p\), the perturbative analysis does not constrain Type IIA and Type IIB in the
same way. In Type IIA, the value of \(\alpha p\), with $\alpha p>2$, determines an allowed interval, whereas in Type IIB the perturbative lower bound coincides with the species-scale upper
bound, as is clear from \eqref{bounds:n:IIA:IIB}. Thus, for Type IIB, fixing \(\alpha p\) fixes \(n\). For Type IIA, instead, fixing
\(\alpha p\) leaves an interval of admissible values of \(n\). This is the same distinction
that appears in the species-scale analysis of Appendix~\ref{app:species:scale:anisotropy}:
Type IIB sits on the species-saturating locus \(\Lambda_{\rm sp}=M_{\rm SUSY}\), whereas
Type IIA can explore the region \(\Lambda_{\rm sp}\geq M_{\rm SUSY}\).

It is instructive to compare our bounds with those obtained in
\cite{Castellano:2021aa}. 
In Type IIA compactifications, allowing $\alpha$ to vary, we find here the common lower bound $n \geq 1/3$, which coincides with
the one found in \cite{Castellano:2021aa} for general $CY_3$ geometries and is independent of the
number $p$ of large dimensions. 
In Type IIB, our analysis
yields stronger lower bounds on $n$ and, most notably, in both Type IIA and Type IIB cases, we are able to increase the lower bounds by lowering the quantity $\alpha p$. Regarding the upper bound, our result follows from imposing the hierarchy
between the scale of SUSY breaking and the species scale, whereas
\cite{Castellano:2021aa} enforces a universal constraint $n \leq 1$.
These correspond to different effective descriptions, depending on whether
the KK–gravitino tower is integrated out or retained.
In the following we will rely on the bound \eqref{strong:n:bound}, and therefore we will allow for values of $n$ bigger than one. However, we find interesting that for the phenomenologically relevant case where $p=2$ with $m_{\text{KK}}^{-1} \sim \mu\mathrm{m}$, our analysis uniquely selects the value $n=1$.

\section{Phenomenological implications from the bounds} \label{chap:finding:scenarios}

Having established quantitative relations between the gravitino mass and the properties
of the extra dimensions, 
we now explore the phenomenologically relevant \textit{scenarios}. In each scenario we analyze the interconnected quantities
$\alpha$, $p$, $n$, $m_{3/2}$, $R \sim 1/m_{\text{KK}}$, $M_{\text{SUSY}}$, and $\Lambda_{\text{sp}}$. 

We remain agnostic about the microscopic origin of $\mathcal{N}=1$ SUSY breaking and assume it
takes place in a hidden sector. The observable phenomenology is then determined by the mediation
mechanism transmitting SUSY breaking to the visible sector. For simplicity we focus on the two dominant possibilities, gauge and gravity mediation,
which are characterized by different hierarchies between the gravitino mass $m_{3/2}$
and the scale $M_{\text{SUSY}}$. 

This leads to two qualitatively distinct regimes that we analyze in the following.
The first corresponds to a light gravitino, gauge mediation, and compactifications
with micron-sized extra dimensions.
The second corresponds instead to a heavier gravitino, gravity mediation,
and a different compactification scale.

\subsection{Scenarios with micron-sized extra dimensions}
Current experimental bounds on deviations from Newton’s law (see Section \ref{sec:bounds:on:extra:dimensions}) require the largest extra dimensions to satisfy
$R \leq \mathcal{O}(10)\, \mu\text{m}$.
Next-generation torsion-balance experiments \cite{Lee:2020aa} are expected to probe distances down to the micron range, therefore, we adopt a reference compactification scale
$R \sim \mu\text{m}$, compatible with the upcoming experimental sensitivity and therefore phenomenologically well motivated.
Astrophysical constraints (table \ref{tab:correct:astrophisical:bounds:on:l}) then restrict the number of large extra dimensions at this scale to $p=1$ or $p=2$. Using $m_{\text{KK}} = R^{-1}$, we can estimate the corresponding species scale $\Lambda_{\rm sp}$, the quantum-gravity cutoff of the theory, through Eq.~\eqref{species:scale:effective}. For the maximally anisotropic case, $\alpha=1$:

\begin{itemize}

    \item \textbf{$p=1$}: $\Lambda_{\rm sp} = \mathcal{O}( 10^9) \,\text{GeV}$. This reproduces the characteristic scale emerging in the Dark Dimension scenario \cite{Montero:2022prj,Anchordoqui:2023oqm} with one large extra-dimension, but here it follows purely from the Gravitino Conjecture and does not depend on the value of the dark-energy density. Remarkably, the scale is close to the cutoff observed in the flux of ultra-high-energy cosmic rays (GZK limit) \cite{Greisen:1966jv, Zatsepin:1966jv, Abraham_2008, Anchordoqui_2019, Montero:2022prj}.
    
    \item \textbf{$p=2$}: $\Lambda_{\rm sp} = \mathcal{O}( 10) \,\text{TeV}$. A fundamental gravitational scale around the TeV region can alleviate the electroweak hierarchy problem through the presence of large extra dimensions at micron size, in the spirit of large-extra-dimension scenarios \cite{Arkani-Hamed:1998jmv}. In this case, supersymmetry breaking naturally occurs at the same scale as $\Lambda_{\rm sp}$. 
\end{itemize}

\noindent
From the relation \eqref{length:gravconj:equality}, at fixed compactification scale \(R\), larger
values of the gravitino mass require larger values of the exponent \(n\)\footnote{In particular,
combining the GC relation with the universal experimental lower bound
\(m_{3/2}\geq 0.1\,\mathrm{eV}\) \eqref{bound:gravitino:mass:LHC} gives, for fixed \(m_{\rm KK}\),
$n \geq \dfrac{\log (m_{\rm KK}/M_\text{P})}{\log (0.1\,\text{eV}/M_\text{P})}$. For $m^{-1}_{\rm KK}\sim \mu m$, this reads $n \geq 1$, where we have set \(\lambda_{3/2}=1\)}. For micron-sized extra dimensions,
\(R\sim \mu{\rm m}\),
this gives
\begin{equation}
\label{alpha<2:lower_n_bound}
    1 \leq n \,.
\end{equation}
This lower bound for $n$ comes from the experimental lower bound on $m_{3/2}$ and is specific for micron-sized extra dimensions.

This observation is especially useful in the Type IIA regime with
\(\alpha p<2\). In this case, the perturbative argument used above does not provide a lower bound on \(n\). However, if we restrict to micron-sized extra dimensions, the
experimental gravitino bound supplies the missing lower bound, and one obtains
\begin{equation}
\label{micron:extra:dim:IIA:n:bound:alpha<2}
    \text{IIA:}\qquad
    1\leq n \leq
    \frac{\alpha p+2}{2\alpha p},
    \qquad
    \alpha p<2,
    \qquad
    R\sim \mu{\rm m}.
\end{equation}

Therefore, in the micron-sized regime and for \(\alpha p<2\), Type IIA and Type IIB are
effectively constrained by the same lower bound \eqref{alpha<2:lower_n_bound}. Together with the upper bound from the species-scale consistency condition \eqref{strong:n:bound}, this yields
the same allowed interval for \(n\) in both theories. At the boundary value
\(\alpha p=2\), the allowed values for $n$ collapse to $n=1$ (see table \ref{tab:allowed:n:fixed:alpha}). After varying $\alpha p$, the numerical intervals of allowed values for $n$ coincide in Type IIA and Type IIB SUGRA. However, the underlying structure at fixed $\alpha p$ is different.

Focusing on the phenomenologically relevant cases \(p=1\) and \(p=2\), one can then
translate these constraints into the corresponding allowed ranges for \(\alpha\). Accordingly, $M_{\rm SUSY}$ scales with $m_{3/2}$. The resulting scenarios with the corresponding parameter space is summarized in table~\ref{tab:scenarios}.

However, by fixing the quantity $\alpha p$, in Type IIA we can explore an interval of $n$ values, instead in Type IIB this uniquely fixes $n$. This influences the relation between $\Lambda_{\rm sp}$ and $M_{\rm SUSY}$, allowing for $\Lambda_{\rm sp}\gg M_{\rm SUSY}$ only in Type IIA, while in Type IIB we have $\Lambda_{\rm sp}= M_{\rm SUSY}$ (see Appendix \ref{app:species:scale:anisotropy}). This difference is made quantitative for $\alpha=1$ in table~\ref{tab:scenarios:alpha=1}.

\begin{table}[!htbp]
\centering
\renewcommand{\arraystretch}{1.5}
\begin{tabular}{|l|c|c|c|}
\cline{3-4}
\multicolumn{2}{c|}{} & \(p=1\) & \(p=2\) \\
\hline
Compactification scale & \(R\)
& \(10^{-6}\,\mathrm{m}\)
& \(10^{-6}\,\mathrm{m}\) \\

Anisotropy parameter & \(\alpha\)
& \(2 \geq \alpha \geq 1\)
& \(1\) \\

Species scale & \(\Lambda_{\rm sp}\)
& \(10^4\,\mathrm{GeV}\leq \Lambda_{\rm sp}\leq 10^9\,\mathrm{GeV}\)
& $10^4\,\mathrm{GeV}$ \\

\multirow{2}{*}{Exponent parameter}
& \multirow{2}{*}{\(n\)} & \textbf{IIA: } \(1\leq n\leq \frac{\alpha p+2}{2\alpha p}\)
&  \multirow{2}{*}{\(1\)} \\

& & \textbf{IIB: } \,\, \(n=\frac{\alpha p+2}{2\alpha p}\)
& \\

Gravitino mass & \(m_{3/2}\)
& \(0.1\,\mathrm{eV}\leq m_{3/2}\leq 0.1\,\mathrm{GeV}\)
& \(0.1\,\mathrm{eV}\) \\

SUSY-breaking scale & \(M_{\rm SUSY}\)
& \(10^4\,\mathrm{GeV}\leq M_{\rm SUSY}\leq 10^9\,\mathrm{GeV}\)
& $10^4\,\mathrm{GeV}$ \\
\hline
\end{tabular}
\caption{Predicted physical scales for one and two micron-sized extra dimensions. All these values are considered up to numerical factors and with $\lambda_{3/2}=1$. For fixed \(\alpha p\), Type IIA admits an interval of
allowed values of \(n\), while Type IIB fixes \(n\) to the species-saturating value.}
\label{tab:scenarios}
\end{table}

\begin{table}[!htbp]
\centering
\small
\renewcommand{\arraystretch}{1.4}
\begin{tabular}{|l|c|c|c|c|}
\cline{3-5}
\multicolumn{2}{c|}{}
& \multicolumn{2}{c|}{$p=1$} 
& $p=2$ \\
\cline{3-5}
\multicolumn{2}{c|}{}
& \textbf{IIA}
& \textbf{IIB}
&  \textbf{IIA/IIB}\\
\hline

Compactification scale 
& $R$
& $10^{-6}\,\mathrm{m}$
& $10^{-6}\,\mathrm{m}$
& $10^{-6}\,\mathrm{m}$ \\

Anisotropy parameter
& $\alpha$
& $1$
& $1$
& $1$ \\

Species scale
& $\Lambda_{\mathrm{sp}}$
& $10^9\,\mathrm{GeV}$
& $10^9\,\mathrm{GeV}$
& $10^4\,\mathrm{GeV}$ \\

Exponent parameter
& $n$
& $1\leq n\leq 3/2$
& $3/2$
& $1$ \\

Gravitino mass
& $m_{3/2}$
& $0.1\,\mathrm{eV}\leq m_{3/2}\leq 0.1\,\mathrm{GeV}$
& $0.1\,\mathrm{GeV}$
& $0.1\,\mathrm{eV}$ \\

SUSY-breaking scale
& $M_{\mathrm{SUSY}}$
& $10^4\,\mathrm{GeV}\leq M_{\mathrm{SUSY}}\leq 10^9\,\mathrm{GeV}$
& $10^9\,\mathrm{GeV}$
& $10^4\,\mathrm{GeV}$ \\

Hierarchy
& 
& $M_{\mathrm{SUSY}}\leq\Lambda_{\mathrm{sp}}$
& $M_{\mathrm{SUSY}}=\Lambda_{\mathrm{sp}}$
& $M_{\mathrm{SUSY}}=\Lambda_{\mathrm{sp}}$ \\

\hline
\end{tabular}

\caption{Physical scales for one and two micron-sized extra dimensions with $\alpha=1$, and $\lambda_{3/2}=1$. Type IIA SUGRA allows for $ M_{\rm SUSY} \ll \Lambda_{\mathrm{sp}}$, instead in Type IIB always holds $ M_{\rm SUSY}=\Lambda_{\mathrm{sp}} $.}
\label{tab:scenarios:alpha=1}
\end{table}

Remarkably, the predicted values of the species scale for $\alpha=1$ coincide with those appearing in the Dark Dimension framework \cite{Montero:2022prj,Anchordoqui:2025nmb}. Here, however, the scale is not assumed but follows from the Gravitino Conjecture combined with experimental bounds. 
Our setup additionally predicts the gravitino mass associated with the allowed values of $n$. In particular, the case $p=1$ with $n\simeq1$ is only admissible in the Type IIA micron-sized regime and gives
\(M_{\rm SUSY}\sim \mathcal{O}(10 \text{--}100)\,\mathrm{TeV}\), numerically similar to the scenario discussed in \cite{Anchordoqui:2023oqm}, although obtained here without invoking a relation to the cosmological constant.

Combining the bounds \eqref{gravity:bounds} and \eqref{gauge:bounds}, we find that the allowed gravitino masses lie entirely in the regime typically associated with gauge mediation. 
One may then ask how the compactification scale would change if the gravitino mass were instead large enough to enter the gravity-mediated regime.

Finally, in the $p=2$ case, we consider values $R\lesssim \mu \mathrm{m}$ that saturate the bounds summarized in table \ref{tab:correct:astrophisical:bounds:on:l}, which are marginally compatible with current constraints. In this regime, the lower bound on
\(m_{3/2}\) translates, through the GC, into values of \(n\) very close
to unity. Therefore, in tables ~\ref{tab:scenarios} and \ref{tab:scenarios:alpha=1} slightly smaller radii than \(R\sim\mu{\rm m}\) are allowed, but the phenomenologically
relevant region remains concentrated around \(n\simeq1\).

\subsection{Scenario with higher gravitino mass}

We now consider the regime in which SUSY breaking is gravity mediated.
From \eqref{gravity:bounds} the transition to gravity mediation occurs for $m_{3/2} \geq  1\,\mathrm{TeV}$, which corresponds to a lower bound on the SUSY breaking scale as $M_{\mathrm{SUSY}} \geq 5\times 10^{10}\,\mathrm{GeV}$.

Using the gravitino conjecture \eqref{GravConjFinal} in inverted form,
the compactification length follows from the gravitino mass. The large value
of $m_{3/2}$ implies a correspondingly large KK scale and therefore very
small extra dimensions,
\begin{equation}
R \ll \mu\mathrm{m}.
\end{equation}
Hence astrophysical constraints are automatically satisfied and no restriction on the number $p$ of large extra dimensions arises.
In this regime, the full allowed intervals of $n$ must be considered,
leading to different predictions for Type IIA and Type IIB compactifications.
However, the associated compactification scales lie far beyond experimental reach, so the distinction has no direct phenomenological impact. Moreover, since the species scale  is related to the effective KK scale through Eq.~\eqref{species:scale:effective},
larger values of $n$, corresponding to larger compactification lengths, lead to smaller values of $\Lambda_{\mathrm{sp}}$. 

The same hierarchies explained in the previous section hold between the species scale and SUSY-breaking scale in Type IIA and Type IIB cases. An exhaustive analysis of this can be found in the Appendix~\ref{app:species:scale:anisotropy}.

All these features of the gravity–mediated scenario are summarized in tables~\ref{tab:IIA:analysis} and~\ref{tab:IIB:analysis} for any value of $\alpha$. In particular, for Type IIA SUGRA, we consider only values $\alpha p>2$, compatible with  our perturbative bound on $n$.

\begin{table}[h!]
\centering
\small
\setlength{\tabcolsep}{6pt}
\renewcommand{\arraystretch}{1.25}

\begin{tabular}{|c |c| c| c|}
\hline
$p$ & allowed $n$ & $R\,[\mathrm{m}]$ & $\Lambda_{\mathrm{sp}}\,[\mathrm{GeV}]$ \\
\hline
1 & $1/3 - 1$   & $1\times10^{-29} - 2\times10^{-19}$ & $5\times10^{10} - 4\times10^{14}$ \\
2 & $1/3 - 1$   & $1\times10^{-29} - 2\times10^{-19}$ & $5\times10^{10} - 4\times10^{14}$ \\
3 & $1/3 - 5/6$ & $1\times10^{-29} - 5\times10^{-22}$ & $5\times10^{10} - 4\times10^{14}$ \\
4 & $1/3 - 3/4$ & $1\times10^{-29} - 3\times10^{-23}$   & $5\times10^{10} - 4\times10^{14}$ \\
5 & $1/3 - 7/10$& $1\times10^{-29} - 5\times10^{-24}$ & $5\times10^{10} - 4\times10^{14}$ \\
6 & $1/3 - 2/3$ & $1\times10^{-29} - 1\times10^{-24}$ & $5\times10^{10} - 4\times10^{14}$ \\
\hline
\end{tabular}

\caption{Predicted ranges of compactification length and species scale in \textbf{Type IIA} compactifications for a gravitino mass $m_{3/2}=\mathrm{TeV}$, corresponding to the gravity--mediated regime ($M_{\mathrm{SUSY}}=5\times10^{10}\,\mathrm{GeV}$). For $p=1$, the validity of our analysis reduces the range of $\alpha$ to $2\le \alpha \le 6$.}
\label{tab:IIA:analysis}
\end{table}

\begin{table}[ht]
\centering
\small
\setlength{\tabcolsep}{6pt}
\renewcommand{\arraystretch}{1.25}

\begin{tabular}{|c| c| c| c|}
\hline
$p$ & allowed $n$ & $R\,[\mathrm{m}]$ & $\Lambda_{\mathrm{sp}}\,[\mathrm{GeV}]$ \\
\hline
1 & $2/3 - 3/2$   & $1\times10^{-24} - 1\times10^{-11}$ & $5\times10^{10}$ \\
2 & $2/3 - 1$   & $1\times10^{-24} - 2\times10^{-19}$ & $5\times10^{10}$ \\
3 & $2/3 - 5/6$ & $1\times10^{-24} - 5\times10^{-22}$ & $5\times10^{10}$ \\
4 & $2/3 - 3/4$ & $1\times10^{-24} - 3\times10^{-23}$   & $5\times10^{10}$ \\
5 & $2/3 - 7/10$& $1\times10^{-24} - 5\times10^{-24}$ & $5\times10^{10}$ \\
6 & $2/3$       & $1\times10^{-24}$                     & $5\times10^{10}$ \\
\hline
\end{tabular}

\caption{Predicted ranges of compactification length and species scale in \textbf{Type IIB} compactifications for a gravitino mass $m_{3/2}=\mathrm{TeV}$, corresponding to the gravity--mediated regime ($M_{\mathrm{SUSY}}=5\times10^{10}\,\text{GeV}$).}
\label{tab:IIB:analysis}
\end{table}

\section{Conclusions}\label{Sec:Conclusions}

In this work we explored the phenomenology emerging from a direct connection between the gravitino mass and the Kaluza--Klein scale in four-dimensional $\mathcal{N}=1$ supergravity arising from Type II compactifications with $p$ large extra dimensions, allowing for an anisotropic internal geometry
parametrized by the scaling exponent $\alpha$. This connection was first suggested in \cite{Cribiori:2021aa,Castellano:2021aa} under the name of Gravitino Conjecture; here we focus on its physical consequences.

We analyzed the anisotropy induced in the compactification manifold by the \(p\) large
directions through the scaling relation $\mathcal V \sim \mathcal V_p^\alpha $, where \(\mathcal V\) is the full six-dimensional internal volume and \(\mathcal V_p\) is the
volume of the large \(p\)-cycle. For fixed \(p\), the anisotropy parameter satisfies
\begin{equation}
    1\leq \alpha \leq \frac{6}{p},
\end{equation}
with \(\alpha=1\) corresponding to maximal anisotropy and \(\alpha=6/p\) to the isotropic
endpoint. Equivalently, the combination \(\alpha p\) measures the effective number of
internal directions participating in the decompactification limit.

We derived the $\alpha$-dependent volume contributions to the Kähler potential and obtained sharp constraints on the scaling exponent $n$ that links $m_{3/2}$ to the size and number of large extra dimensions. Interestingly, we find that the linear relation $n=1$, which would correspond 
to the simplest proportionality between $m_{3/2}$ and $m_{\rm KK}$, is only compatible with $\alpha p \leq 2$. This would provide an intriguing theoretical explanation for why only one or two large extra dimensions remain experimentally viable at the micron scale.

Our perturbative procedure treats Type IIA and Type IIB differently and fixing \(\alpha p\) leaves a range of possible \(n\)-values in Type IIA, but selects
a single value in Type IIB. This difference is reflected in the species scale: Type IIB
saturates \(\Lambda_{\rm sp}=M_{\rm SUSY}\), while Type IIA can realize
\(M_{\rm SUSY}\ll\Lambda_{\rm sp}\).

We studied two qualitatively distinct classes of scenarios:
\begin{enumerate}
\item \emph{Micron-sized extra dimensions.}
For $R = 10^{-6}\,\mathrm{m}$ only $p=1,2$ are allowed. The experimental bound on the gravitino mass implies \(n\geq1\) and, for the maximally anisotropic case
\(\alpha=1\), the two theories behave differently when \(p=1\). The gravitino mass lies in the eV--GeV range and supersymmetry breaking is
necessarily gauge mediated. 
These scenarios are experimentally testable in upcoming short-distance gravity
experiments and predict correlated values of $m_{3/2}$, $M_{\mathrm{SUSY}}$,
and the compactification scale. 
Interestingly, the case with two large extra dimensions implies a low quantum
gravity scale and a SUSY breaking scale both of order $\mathcal{O}(10)\,\mathrm{TeV}$.

\item \emph{Heavy gravitino regime.}
For $m_{3/2}= \mathrm{TeV}$ one obtains gravity-mediated
supersymmetry breaking with $M_{\mathrm{SUSY}}\sim10^{10}\,\mathrm{GeV}$
and extremely small values for the compactification radius $R$, far beyond experimental reach.
Although less accessible, these scenarios complete the map connecting SUSY
breaking to the geometry of extra dimensions.
\end{enumerate}

Crucially, in our framework the tower of states is directly controlled by supersymmetry breaking rather than by an independent infrared scale. This distinguishes it from the Dark Dimension scenario \cite{Montero:2022prj, Anchordoqui:2025nmb}, where the KK scale is set independently of the gravitino mass and the SUSY breaking scale. In the Dark Dimension scenario, the KK tower is tied to the observed dark energy density $\Lambda$, and no quantitative predictions for $m_{3/2}$ 
or $M_{\mathrm{SUSY}}$ can be made. 
Here, by contrast, measurements on either the particle physics side or the gravitational side immediately constrain the other,
leading to a predictive bridge between the two sectors. 

Let us emphasize that our results hold for a proportionality parameter $\lambda_{3/2}$ in Eq.~\eqref{GravConjFinal} of order unity (in our analysis we fixed $\lambda_{3/2}=1$). 
In the Dark Dimension scenario, instead, the analogous parameter relating the KK mass to the dark--energy scale must be tuned to $\lambda\sim\mathcal{O}(10^{-3})$ in order to obtain extra--dimension sizes compatible with current experimental bounds.  
Equivalently, the universal lower bound on the gravitino mass $m_{3/2}\gtrsim 0.1\,\mathrm{eV}$ automatically translates into an upper bound on the compactification length $R\lesssim\mu\mathrm{m}$, without any adjustment of parameters, as we already noticed in the introduction of this work. 
By contrast, the intrinsic length scale $\Lambda^{-1/4}$ associated with dark energy is of order the millimeter and therefore requires a small tuning to evade present short--distance gravity constraints.

Overall, our analysis provides a predictive bridge between the geometry of extra dimensions and supersymmetry breaking, translating measurements of gravitational short--distance scales into information on particle--physics parameters, and vice versa.
Some of the consequences of this framework, including its implication for dark energy, will be explored in a forthcoming work~\cite{Bersigotti:workinprogress}.

\section*{Acknowledgments}
We are indebted to Luis Anchordoqui, Ivano Basile, Alessandro Borys, Gonzalo Casas, Bernardo Fraiman, Álvaro Herráez, Joaquín Masias, Carmine Montella, Georgina Staudt, and Matteo Zatti for valuable suggestions and comments on previous versions the draft. We also thank Arthur Hebecker and Gia Dvali for enlightening discussions. The work of D.L. is supported by  the German–Israel Project (DIP) on Holography and the Swampland. L.B. thanks Northeastern University and the Corfu Summer Institute for the stimulating environments provided during, respectively, the String Pheno ’25 conference and the Workshop on Quantum Gravity and Strings, where parts of this work were presented.  M.S. acknowledges the support of the University of Catania through PIAno di inCEntivi per la RIcerca di Ateneo 2024/2026 - Project "COSMOgraM".

\appendix

\section{Spectrum and \texorpdfstring{$\mathcal{N}=1$}{N=1} action for Type II theories}\label{AppII}

In this appendix we will clarify our conventions in denoting quantities in the spectrum and the tree-level effective action in $\mathcal{N}=1$ SUGRA form of both Type IIA compactified in an orientifold with $O6$-planes and Type IIB compactified in orientifolds with $O3/O7$- or $O5/O9$-planes. Then, we will provide also the mirror symmetry results for an $\mathcal{N}=1$ version of the moduli space which we recall in the main text.

In $\mathcal{N}=1$ SUGRA the action is expressed in terms of a Kähler potential $K$, an holomorphic superpotential $W$ and the holomorphic gauge-kinetic coupling functions $f$, as follows
\be\label{EffectiveSUGRAActionGeneral}
S^{(4)} = - \int \frac{1}{2} R \ast 1 
+ K_{i\bar{j}}\, d\phi^i \wedge \ast d\bar{\phi}^{\bar{j}} 
+ \frac{1}{2} \text{Re}\, f_{ab}\, F^a\wedge \ast F^b
+ \frac{1}{2} \text{Im}\, f_{ab}\, F^a \wedge F^b
+ V \ast 1 ,
\ee
where $\phi^i$ collectively denote all the complex scalars in the theory, $K_{i\bar{j}}$ is the Kähler metric satisfying $K_{i\bar{j}}=\partial_i \bar{\partial}_{\bar{j}}K$, and the scalar potential is given by the equation \ref{scalar:potential}. Moreover, the general situation we consider is a Calabi-Yau orientifold constructed from a Calabi-Yau manifold by modding out a discrete symmetry 
\be
\label{Orientifold}
\mathcal{O}=\Omega_p(-1)^{F_L}\sigma\,,
\ee
which include the world-sheet parity $\Omega_p$, the fermion number of the left-moving sector $(-1)^{F_L}$ and an involution in the Calabi-Yau manifold $Y_6$ alone, so that $\sigma^2=1$. In Type IIA case, if one wants to preserve $\mathcal{N}=1$ SUSY, then the involution $\sigma$ has to be anti-holomorphic. In the following, we will denote with the hat the ten-dimensional quantities.

Following \cite{grimm_effective_2004}, Type IIA orientifolds with anti-holomorphic involutions generically admit $O6$ planes. The space of harmonic forms splits under the
involution $\sigma$ into even and odd eigenspaces 
\be\label{HarmonicformsSplitted}
H^p(Y_6)=H^p_+\oplus H^p_-
\ee
and the basis elements of the non-trivial cohomology groups are $\omega_{\alpha_+}\,,\omega_{\alpha_-}$ denoting even and odd $(1,1)$-forms, $\tilde{\omega}^{\alpha_+}\,,\tilde{\omega}^{\alpha_-}$ denoting even and odd $(2,2)$-forms, where also hold the relations $h^{(1,1)}_+=h^{(2,2)}_-\,,h^{(1,1)}_-=h^{(2,2)}_+$ between the numbers of the forms. Furthermore, we have $h^{(3,3)}_+=0\,,h^{(3,3)}_-=1$ and $h^{(0,0)}_+=1\,,h^{(0,0)}_-=0$. Then, the forms of the full theory can be expanded in terms of convenient basis of harmonic forms \footnote{In the RR sector of the $\mathcal{N}=2$ theory, there is also the $A_1$ form, which however is fully projected out by orbifolding the Calabi-Yau.} as:
\begin{align}
\label{TheoryExpansions}
    J=v^{\alpha_-}\omega_{\alpha_-}, && \hat{B}_2 = b^{\alpha_-}\omega_{\alpha_-}, && {\alpha_-}=1,\dots,h^{(1,1)}_-\,,\\ \nonumber
    \hat{C}_3 = c_3 + A^{\alpha_+}\wedge\omega_{\alpha_+} + C_3, && C_3 = \xi^\kappa \alpha_\kappa, && \alpha_\kappa \in H^{(3)}_+\,,
\end{align}
where $\xi^\kappa$ are $h^{(2,1)}+1$ real scalars, $A^{\alpha}$ are $h^{(1,1)}_+$ one-forms and $c_3$ is a three-form in four-dimensions with no physical degrees of freedom, corresponding just to a constant flux parameter in the superpotential. Furthermore, $v^{\alpha_-}$ and $b^{\alpha_-}$ are space-time scalars and we can combine them, as in $\mathcal{N}=2$ theory, into complex coordinates 
\be\label{IIACoordinates}
t^{\alpha_-} = v^{\alpha_-} + ib^{\alpha_-}\,.
\ee
The effect of the orientifold is therefore the one of reducing by a half the dimension of the Kähler moduli space and similarly of the deformations of the complex structure, which in turn remains the same as in the $\mathcal{N}=2$ case. The appropriate complex fields arise from the complexified three-form
\begin{equation}
    \Omega_c = C_3 + 2i\textrm{Re}(C\Omega)\,,
\end{equation}
where $\Omega$ is the holomorphic three-form, only defined up to complex rescalings by a holomorphic function $e^{-h(z)}$ and which, under the further action of $\sigma$ changes by a (real) rescaling $\Omega\rightarrow\Omega e^{-\text{Re}(h)}$. To compensate for this, a complex compensator $C$ is introduced which transforms as $C\rightarrow C e^{+\text{Re}(h)}$ and make the function $C\Omega$ invariant. Now, $\Omega_c$ can be expanded in a basis $\alpha_\kappa$ of $H^{(3)}_+$, with $\kappa=0, \dots, h^{(2,1)}$, as 
\[\Omega_c=2N^\kappa \alpha_\kappa\,,\]
where the moduli $N^\kappa$ are given by $\xi^\kappa$ that combines with $h^{(2,1)}$ real complex structure deformations to form chiral multiplets. Let us summarize the resulting $\mathcal{N} = 1$ spectrum. It assembles into a gravitational multiplet, $h^{(1,1)}_+$ vector multiplets with bosonic components $A^{\alpha_+}$, $h^{(1,1)}_-$ chiral multiplets with bosonic components $t^{\alpha_-}$ and  $h^{(2,1)}+1$ chiral multiplets with bosonic components $N^\kappa$.

The moduli space has the product structure 
\begin{equation}
\label{typeIIA:orientifold:product:moduli:space}
    \mathcal{M}^{h^{(1,1)}_-}_K \times \mathcal{M}^{h^{(2,1)}+1}_Q,
\end{equation}
where $\tilde{\mathcal{M}}^K$ is a subspace of the $N=2$ moduli space ${\mathcal{M}}^K$ with dimension $h^{(1,1)}_-$ spanned by the complex Kähler deformations $t^a$, instead $\tilde{\mathcal{M}}^Q$ is a subspace of the quaternionic manifold ${\mathcal{M}}^Q$ with dimension $h^{(2,1)}+1$ and spanned by the complex structure deformations $q^K$, the four-dimensional dilaton $\phi_{(4)}$ and the scalars $\xi^K$ arising from $C_3$.

The metric of $\tilde{\mathcal{M}}^K$ is a simple truncation of the $N=2$ one and remains special Kähler. The Kähler potential for this component is given by 
\begin{equation}
\label{Kähler:potential:Kähler:component:IIA:orbifold}
    K^K = -\ln \left[ \frac{i}{6} \mathcal{K}_{abc}(t-\bar{t})^{a}(t-\bar{t})^{b}(t-\bar{t})^{c}\right] = -\ln \left[\frac{4}{3}\int_Y J\wedge J\wedge J\right]\,.
\end{equation}
It is known that $K^K$ obeys the standard no-scale cancellation \cite{Cremmer:1983bf} \[ K_{t^a}K^{t^a\bar{t}^b}K_{\bar{t}^b}=3.\] $J$ is the Kähler form in string frame and we indicated the topological intersection numbers as
\begin{align}
\label{intersections}
    \mathcal{K}_{abc} = \int_Y \omega_a \wedge \omega_b \wedge \omega_c, && \mathcal{K}_{ab} = \int_Y \omega_a \wedge \omega_b \wedge J= \mathcal{K}_{abc}v^c, \\ \nonumber \mathcal{K}_{a} = \int_Y \omega_a \wedge J \wedge J = \mathcal{K}_{abc}v^bv^c, && \mathcal{K} = \int_Y J \wedge J \wedge J = \mathcal{K}_{abc}v^av^bv^c\,,
\end{align}
in terms of the real moduli $v^a$ of equation \eqref{TheoryExpansions}. With this conventions, the volume $\mathcal{V}$ of the Calabi-Yau manifold in string frame is given in terms of the two-cycle moduli as
\begin{equation}
    \label{volume:Calabi:Yau}
    \mathcal{V} = \frac{\mathcal{K}}{6}=\frac{1}{6}\mathcal{K}_{abc}v^av^bv^c\,.
\end{equation}
The geometry of the second component $\tilde{\mathcal{M}}^Q$ is instead considerably more involved, but for what interests us, we can derive the integral representation of the Kähler potential for this component as
\begin{equation}
\label{Kähler:potential:Quaternionic:component:IIA:orbifold}
    K^Q = -2\ln \left[ 2 \int_Y \textrm{Re}(C\Omega) \wedge \ast \textrm{Re}(C\Omega)\right] = -\ln e^{-4\phi_{(4)}} = 4\phi_{(4)}\,,
\end{equation}
where the four dimensional dilaton $\phi_{(4)}$ can be defined in terms of the ten-dimensional one $\phi_{(10)}$ according to 
\begin{equation}
    \label{four:dimensional:dilaton:definition}
    e^{\phi_{(4)}} = e^{\phi_{(10)}}(\mathcal{K}/6)^{-\frac{1}{2}}.
\end{equation}

Let us summarize the results obtained so far. We found that the moduli space of $\mathcal{N}=1$ orientifolds is indeed the product of two Kähler spaces with total Kähler potential 
\begin{align}
\label{total:kähler:potential:IIA:orbifold}
K = K^K + K^Q, \\ \nonumber K^K = -\ln \left[\frac{4}{3}\int_Y J\wedge J\wedge J\right], && K^Q = -2\ln \left[ 2 \int_Y \textrm{Re}(C\Omega) \wedge \ast \textrm{Re}(C\Omega)\right].
\end{align}
The VEVs for the moduli can be fixed by turning on fluxes to generate a potential. However, we will not review this part and we explain in section \ref{subsubsect:GravitinoMassAndVolumeDependence} how to handle the superpotential $W$ for what interests us.

Now, following \cite{Grimm:2005aa} we impose the projection \eqref{Orientifold} on the Type IIB case which now requires $\sigma$ to be a holomorphic involution. The space of the harmonic forms splits again as \eqref{HarmonicformsSplitted}, where now $h^{(1,1)}_{\pm}=h^{(2,2)}_{\pm}\,,\,h^{(2,1)}_{\pm}=h^{(1,2)}_{\pm}$ and $h^{(3,0)}_{+}=h^{(0,3)}_{+}=0\,,\,h^{(3,0)}_{-}=h^{(0,3)}_{-}=1$ for $O3/O7$ orientifolds, instead $h^{(3,0)}_{+}=h^{(0,3)}_{+}=1\,,\,h^{(3,0)}_{-}=h^{(0,3)}_{-}=0$ $O5/O9$ orientifolds. Furthermore, we have $h^{(0,0)}_{+}=h^{(3,3)}_{+}=1\,,\,h^{(0,0)}_{-}=h^{(3,3)}_{-}=0$. The basis of harmonic forms are denoted accordingly to the Type IIA case, but considering now the above given numbers for the dimension of the space where they live. We now specify on the $O3/O7$ case, for which we have the following expansion for the $N=1$ spectrum by compactifying in Calabi-Yau orientifold with O3/O7-planes:
\label{deformations:typeIIB:orientifold:O3/O7}
\begin{align}
    &\hat{\phi}=\phi\,, \nonumber \\J&=v^{\alpha_+}\, \omega_{\alpha_+}\,, \,\,\, \hat{B}_2 = b^{\alpha_-}\,\omega_{\alpha_-}\,, \,\,\, \hat{C_0}=C_0\,,\,\,\,\hat{C}_2 = c^{\alpha_-}\,\omega_{\alpha_-}\,, \\ \nonumber \hat{C}_4 &= D^{\alpha_+}_2\wedge \omega_{\alpha_+}+V^{\kappa} \wedge \alpha_{\kappa} +U_\kappa \wedge \beta^\kappa+ \rho_{\alpha_+}\, \tilde{\omega}^{\alpha_+}\,,
\end{align}
where $(\alpha_\kappa, \beta^\kappa)$ is a real symplectic basis of $H^{(3)}_+=H^{(1,2)}_+\oplus H^{(2,1)}_+$, and where both the axion $\hat{C}_0$ and the scalar $\phi$ remain in the spectrum and we denote the corresponding four-dimensional field by ${C}_0$ and $\phi$. Altogether the resulting $\mathcal{N} = 1$ fields for the two setups assembles into a gravitational multiplet, $h^{(2,1)}_\pm$ vector multiplets and $(h^{(2,1)}_\mp + h^{(1,1)} + 1)$ chiral multiplets. 

Given this, and rearranging the given bosonic components in proper Kähler coordinates $(S,T,G)$ where $S$ is the axio-dilaton and the others are combination of moduli, the total Kähler potential can be written as
\begin{align}
\label{total:kähler:potential:IIB:orbifold}
    K = K_{cs}(z,\bar{z})+K_k(S,T,G), \\ \nonumber K_{cs} = -\ln \left[-i \int \Omega(z) \wedge \bar{\Omega}(\bar{z})\right], && \nonumber K_k = -\ln \left[-i(S-\bar{S})\right] - 2\ln \left[\frac{1}{6}\mathcal{K}(S, T, G)\right].
\end{align}
In this case $\mathcal{K} \equiv \mathcal{K}_{\alpha \beta \gamma}v^{\alpha}v^{\beta}v^{\gamma} = 6 \textrm{Vol}(Y)$ is an implicit function of the coordinates $(S,T,G)$, therefore also the relation between $v^{\alpha}$ and $(S,T,G)$ remain implicit. 

Crucially, we are interested in measuring the same effective quantities at four-dimensional level both in Type IIA and in Type IIB, i.e. the four-dimensional dilaton 
\[ \phi^{IIA}_{(4)} = \phi^{IIB}_{(4)} = \phi_{(4)}\,.\] This selects a specific set of Kähler coordinates (using $\alpha_+=1,\dots,h^{(1,1)}_+$ and $\beta-\,,\,\gamma-=1,\dots,h^{(1,1)}_-$
\begin{align}
\label{proper:Kähler:coordinates:string:frame}
    &S = C_0 + ie^{-\phi_{(10)}}\,,\,\,\,\,\,\,\,\,\,\,\,\,\, G^{\beta_-} = c^{\beta-} - S b^{\beta-}\,,& \\ \nonumber &T_{\alpha_+} = \tau_{\alpha_+}+ i\rho_{{\alpha_+}} + \frac{1}{2(S+\bar{S})}\mathcal{K}_{{\alpha_+} \beta_- \gamma_-}G^{\beta_-}(G-\bar{G})^{\gamma_-}\,, 
\end{align}
where $\mathcal{K}_{\alpha_+ \beta_- \gamma_-}$ are intersection numbers as given in \eqref{intersections}, $C_0$ is the RR scalar, $\tau_{\alpha_+}$ is an implicit function of the moduli $v^{\alpha_+}$ and $e^{\phi_{(10)}}$ is the ten-dimensional Type IIB string coupling. Considering orientifold projections such as $h^{(1,1)}_-=0$, hence $h^{(1,1)}_+=h^{(1,1)}$ (i.e. all the two-form moduli $b^{\alpha_-}$ and $c^{\alpha_-}$ are projected out), the Kähler moduli simplify to
\be\label{MinimalKählerCoordinates}
T_\alpha = \tau_\alpha + i\rho_\alpha\,,
\ee
where now $\alpha=1,\dots,h^{(1,1)}$. The real part of the Kähler moduli measures 4-cycle volumes and its related with a homogeneous quadratic function to the quantities $v^{\alpha_+}$
\be\label{relationFourcylesTwocycles}
\tau_{\alpha}=\frac{1}{2}\int_{Y_6} \omega_{\alpha} \wedge J \wedge J = \frac{1}{2}\mathcal{K}_{\alpha \beta \gamma} v^{\beta} v^{\gamma}\,.
\ee

The Kähler potential part which is influenced by this change, now is given by 
\begin{equation}
    \label{ghj}
    K_k(S, G,T) = -2 \ln \left[e^{-2 \phi_{(10)}} \int J \wedge J \wedge J\right] = -\ln (e^{-4\phi_{(4)}}).
\end{equation}
The moduli space has the form
\begin{equation}
\label{typeIIB:orientifold:product:moduli:space}
    \mathcal{M}^{h^{(1,2)}_-}_{cs} \times \mathcal{M}^{h^{(1,1)}+1}_{Q},
\end{equation}
where each factor is a Kähler manifold. In detail, $\mathcal{M}^{h^{(1,2)}_-}_{cs}$ is spanned by the complex structure deformations of the Calabi-Yau, instead $\mathcal{M}^{h^{(1,1)}+1}_{Q}$ includes the Type IIB dilaton and the Kähler deformations of the Calabi-Yau.

Finally, we can briefly analyze the situation of Type IIB orientifold compactification with O5/O9-planes. Also in this case, the Kähler potential one obtains from this orientifolding procedure, can be rewritten as \eqref{total:kähler:potential:IIB:orbifold} when expressed in terms of the variables $\phi$ and $v^{\alpha}$. Indeed, the NS sector variables are the same in both the orientifolding compactifications and differences in the two Kähler potentials only arise when one is considering different Kähler coordinates for each case.
\subsection{Mirror symmetry results}\label{sec:mirror:symmetry}

In this subsection, we report also a few important results regarding mirror symmetry for Calabi-Yau orientifolds from the point of view of the effective action derived in the large volume limit. Again, for details refer to \cite{louis_generalized_2005}. 

For what interests us, the effective actions on both kinds of the Type IIB orientifold procedures have only been computed in the large volume limit, for this we can expect to find agreement only if we also take the large complex structure limit exactly as in the $N=2$ symmetry. However, we will see that the full $N=1$ moduli space is a lot more complicated than the underlying $N=2$ space \cite{Brunner_2004}. Therefore we will firstly inspect the simpler situation of the special Kähler sectors and then the quaternionic one.

\subsubsection*{Mirror symmetry in the special Kähler sectors}

For Type IIA the special Kähler manifold $\tilde{\mathcal{M}}^K_{IIA}$ is spanned by the complexified Kähler deformations $t^a$. Under the mirror symmetry these moduli are mapped to the complex structure deformations and in both cases the Kähler potential is a truncated version of the $N=2$ Kähler potential. One has:
\begin{equation}
    K^K_{IIA} = -\ln \left[\frac{4}{3}\int_Y J\wedge J \wedge J \right] \leftrightarrow K^{cs}_{IIB} = -\ln \left[-i \int \Omega \wedge \bar{\Omega}\right].
\end{equation} 

\subsubsection*{Mirror symmetry in the quaternionic sectors}

What we can say about the Kähler manifolds $\tilde{\mathcal{M}}^{Q}_{IIA}$ and $\tilde{\mathcal{M}}^{Q}_{IIB}$ arising in the reduction of the quaternionic spaces, is that they behave very non trivially under mirror symmetry. Essentially, we have to recall that in \eqref{total:kähler:potential:IIA:orbifold} we defined complex structure coordinates in terms of the $C\Omega$ combination and, at this point, the important observation is that there seem to be two physically inequivalent ways to fix this scale invariance. This affects also mirror symmetry, not allowing for a universal correspondence between IIA and IIB quantities, but instead we have something more similar to two independent correspondences between IIA and IIB with O3/O7 planes and IIA and IIB with O5/O9 planes \cite{Grimm:2005aa}. Without delving much into details, the interesting thing here is that we can always find relations for the dilaton such that
\begin{equation}
    e^{\phi^{IIA}_{(4)}} = e^{\phi^{IIB}_{(4)}}\,,
\end{equation}
where these are the four-dimensional dilatons of the Type IIA and Type IIB theory. This implies that the Kähler potentials \eqref{ghj} and \eqref{Kähler:potential:Quaternionic:component:IIA:orbifold} coincide in the large volume - large complex structure limit.

\section{Conventions for Kaluza--Klein compactifications}
\label{sec:m_kk:compactification}

We want now to recall the relation between $m_{\text{KK}}$ and the volume of the internal manifold, explaining our conventions for the relation between Einstein and String frames, and considering how the formulas change when the volume is given by $\mathcal{V}_p$ and when the anisotropy parameter $\alpha$ is introduced.

From the gravitational part of the $(d+ p)$-dimensional action of $\mathcal{N}=1$ SUGRA, we have, upon dimensionally reducing up to $d$ dimensions, that
\begin{align}
    \label{action:comparison}
    S_{\text{grav}} \supset {M^{p+d-2}_s}\int d^{p+d}x \sqrt{-G^{(S)}} e^{-2\phi}G^{(S)MN}\mathcal{R}^{(S)}_{MN} \\ \sim \underbrace{\frac{M^{p+d-2}_s}{g_{(p+d)}^2}{{V}_p}^{(E)}}_{M^2_{P,(p+d)}}\int d^{d}x \sqrt{-g^{(E)}} g^{(E)\mu \nu} R^{(E)}_{\mu \nu} \label{second} \\ \sim \underbrace{\frac{M^{d-2}_s}{g^2_{(d)}}}_{M^2_{P,(d)}}\int d^{d}x \sqrt{-g^{(E)}} g^{(E)\mu \nu} R^{(E)}_{\mu \nu}\,.\label{third}
\end{align}
Here $(S)$ and $(E)$ are the String frame and the Einstein frame, instead $G_{MN}$, $\mathcal{R}_{MN}$, $g_{(p+d)}^2$ and $g_{\mu \nu}$, $R_{\mu \nu}$, $g^2_{(d)}$ are respectively the $(d+p)$-dimensional and $d$-dimensional metric, Ricci tensor and string coupling. From now on we will be more precise in denoting the different Planck masses considered, using $M_{\text{P},(p+d)}$ to denote the $(d+p)$-dimensional Planck mass and $M_{\text{P},(d)}$ the $d$-dimensional one. Moreover, $M_s\simeq \ell_s^{-1}$ is the string scale and $\mathcal{V}_{p}={V}^{(E)}_p / \ell^p_s$ is the volume of the internal $p$-cycle in string units and in Einstein frame. Throughout this work we
neglect warping effects and drop numerical factors of order $2\pi$. The reduced
four-dimensional Planck mass is denoted by
\begin{equation}
    M_{\rm P}\equiv M_{{\rm P},(4)} = \sqrt{1/8\pi G_N}
    \simeq 2.48\times 10^{18}\,{\rm GeV}\,.
\end{equation}

The passage from \eqref{second} and \eqref{third} tells us about the well-known relation between the string couplings, that for a general compactification to a $d$-dimensional theory is
\begin{equation}
\label{dilatons:without:alpha}
    g^2_{(d)} = \frac{g^2_{(p+d)}}{\mathcal{V}^{\frac{2}{d-2}}}\,.
\end{equation}
This equation, which is found considering the large volume limit, is also considered to be the \textit{definition} of the $d$-dimensional string coupling. Fixing ourselves in $d=4$ and considering in this case $\alpha =1$, the equation \eqref{dilatons} can be equivalently rewritten using \eqref{Scaling:Alpha}, in terms of the $p$-cycle volume and $\alpha$, as 
\begin{equation}
\label{dilatons}
    g^2_{(4)} = \frac{g^2_{( p+4)}}{\mathcal{V}_p^{\alpha}}\,.
\end{equation}
The equation \eqref{dilatons} is valid since we consider the remaining $6- p$ dimensions to be compactified at very small scales, thus
\begin{equation}
    g_{( p+4)} \sim g_s\,,
\end{equation}
since the volume of the small dimensions in string units is almost one. We will then use $g_s$ in place of $g_{(p+4)}$ in all the following computations. Remember also that we use the string couplings $g$ in place of the dilaton $\phi$ in our computations, where the relation between the two is 
\begin{equation}
\label{conversion:dilaton:coupling}
    g = e^{\langle \phi \rangle},
\end{equation}
where the $\langle \cdot \rangle$ denotes the VEV of the dilaton field.

\medskip
We now specialize to compactifications in which \(p\) internal directions are
parametrically larger than the remaining \(6-p\) directions, hence $\alpha=1$. The volume of the large
\(p\)-cycle, measured in string units, is denoted by \(\mathcal V_p\). If the corresponding
common radius is \(R\), then
\begin{equation}
\label{Vp:KK:definition}
    \mathcal V_p=(R M_s)^p \,.
\end{equation}
The lightest KK scale associated with
the large directions is therefore
\begin{equation}
\label{mkk:and:volume}
    m_\text{KK,1}
    =
    \frac{1}{R}
    =
    \frac{M_s}{\mathcal V_p^{1/p}} .
\end{equation}

However, in an anisotropic compactification where $\alpha>1$, this is not the only tower relevant for the
species count. The remaining \(6-p\) directions also generate KK modes, with a gap
controlled by their radius \(r\). The full six-dimensional internal volume is parametrized
as $ \mathcal V\sim \mathcal V_p^\alpha$ (Eq. \eqref{Scaling:Alpha}),
so that, using
\begin{equation}
    \mathcal V
    =
    (R M_s)^p(rM_s)^{6-p},
\end{equation}
one obtains
\begin{equation}
\label{r:R:relation:conventions}
    r M_s
    =
    (R M_s)^{\frac{p(\alpha-1)}{6-p}} .
\end{equation}
The second KK gap is therefore
\begin{equation}
\label{mr:definition}
    m_\text{KK,2}
    =
    \frac{1}{r}
    =
    \frac{M_s}{(R M_s)^{\frac{p(\alpha-1)}{6-p}}}.
\end{equation}

Thus, the anisotropic compactification contains two KK towers: $m_\text{KK,1}$ and $m_\text{KK,2}$, associated, respectively, to the $p$ large and the $6-p$ transverse directions.  

These towers can be packaged into an effective KK
tower with
\begin{equation}
\label{effective:tower:example}
    m_{\rm KK,eff}
    \equiv
    \left(m_\text{KK,1}^p m_\text{KK,2}^{6-p}\right)^{1/6}
    =
    \frac{M_s}{(R M_s)^{\alpha p/6}} .
\end{equation}
Equivalently, the anisotropy parameter $\alpha$ makes the system behave as if the tower were controlled by an effective number of large directions
\begin{equation}
\label{effective:number:large:dimensions}
    p_{\rm eff}=\alpha p\,.
\end{equation}

Let us also record the corresponding scaling in four-dimensional Planck units. From
\begin{equation}
    M_s \sim g_{(4)}M_{\rm P}\,,
\end{equation}
and from Eq. \eqref{dilatons}, one finds
\begin{align}
\label{KK:scales:IIA:IIB}
    \text{IIA:}\qquad
    m_\text{KK,1}
    &\sim
    g_{(4)}\,\mathcal V_p^{-1/p}M_{\rm P},
    &
    m_{\rm KK,eff}
    &\sim
    g_{(4)}\,\mathcal V_p^{-\alpha/6}M_{\rm P},
    \\[4pt]
    \text{IIB:}\qquad
    m_\text{KK,1}
    &\sim
    g_s\,\mathcal V_p^{-\left(\frac{\alpha}{2}+\frac{1}{p}\right)}
    M_{\rm P},
    &
    m_{\rm KK,eff}
    &\sim
    g_s\,\mathcal V_p^{-\left(\frac{\alpha}{2}+\frac{\alpha}{6}\right)}
    M_{\rm P}.
\end{align}
Here the difference between Type IIA and Type IIB reflects the choice of perturbative
coupling used to describe the large-volume regime: \(g_{(4)}\) in Type IIA and \(g_s\) in
Type IIB, as explained in Section \ref{subsect:dilaton}.

\section{Species scale and anisotropy}
\label{app:species:scale:anisotropy}

Having fixed the Kaluza--Klein conventions in Section~\ref{sec:m_kk:compactification},
we now compute the corresponding species scale\cite{Dvali_2008,Dvali_2010,arkanihamed2005predictive} in the presence of non-trivial anisotropy. This latter, traveling the moduli space in the asymptotic regime, can be understood as the quantum gravity cutoff at which gravity becomes strongly coupled and no EFT description remains valid. The relevant
point is that, although the lightest KK gap is set by the largest radius, $m_{\rm KK,1}=1/R$, the species count is controlled by the effective number of dimensions participating in the
decompactification\cite{Castellano:2021mmx,Castellano:2022bvr,Calderon-Infante:2025ldq}.

\medskip

Denoting by $N$ the number of light particles (or species) in a given theory, the four-dimensional
species scale is defined by
\begin{equation}
\label{species:definition}
    \Lambda_{\rm sp}
    =
    \frac{M_{\rm P}}{\sqrt{N}}\,.
\end{equation}
For several
multiplicative towers,
\begin{equation}
    N
    =
    \prod_i N_i,
    \qquad
    N_i
    =
    \left(
    \frac{\Lambda_{\rm sp}}{m_i}
    \right)^{p_i},
\end{equation}
where \(m_i\) is the mass gap of the \(i\)-th tower and \(p_i\) counts the associated number
of decompactifying directions. Defining
\begin{equation}
    p_{\rm eff}=\sum_i p_i,
    \qquad
    m_{\rm KK,eff}^{p_{\rm eff}}
    =
    \prod_i m_i^{p_i},
\end{equation}
one finds
\begin{equation}
\label{general:formula:lambda:sp}
    \Lambda_{\rm sp}
    =
    m_{\rm KK,eff}^{\frac{p_{\rm eff}}{2+p_{\rm eff}}}
    M_{\rm P}^{\frac{2}{2+p_{\rm eff}}}.
\end{equation}

In the present case there are two KK towers: one associated with the \(p\) large
directions of radius \(R\), and another associated with the remaining \(6-p\) directions
of radius \(r\). The anisotropy relations worked out on Appendix \ref{sec:m_kk:compactification}, imply the effective mass gap \eqref{effective:tower:example}, which in species units reads
\begin{equation}
\label{effective:species:units}
    m_{\rm KK,eff}
    =
    \frac{\Lambda_\text{sp}}{(R \Lambda_\text{sp})^{\alpha p/6}}\,.
\end{equation}
Substituting \eqref{effective:species:units} into
\eqref{general:formula:lambda:sp}, one obtains
\begin{equation}
\label{species:scale:effective}
\boxed{
    \Lambda_{\rm sp}
    =
    R^{-\frac{\alpha p}{2+\alpha p}}
    M_{\rm P}^{\frac{2}{2+\alpha p}}\,.
}
\end{equation}

This expression has the expected limiting behaviours (using $m_\text{KK,1}\simeq R^{-1}$):
\begin{itemize}
    \item For \(\alpha=1\), one recovers the
species scale of a maximally anisotropic compactification with \(p\) large dimensions,
\begin{equation}
    \Lambda_{\rm sp}
    =
    m_{\rm KK}^{\frac{p}{2+p}}
    M_{\rm P}^{\frac{2}{2+p}} .
\end{equation}

\item For \(\alpha=6/p\), one obtains the isotropic six-dimensional result,
\begin{equation}
    \Lambda_{\rm sp}
    =
    m_{\rm KK}^{3/4} M_{\rm P}^{1/4}.
\end{equation}
\end{itemize}

Although \eqref{species:scale:effective} is derived from two KK towers, associated with radii \(R\) and \(r\), its final scaling is equivalent to that of a single effective KK tower associated with \(p_{\rm eff}=\alpha p\) dimensions compactified at the scale \(R\).

\medskip
In our setup, we consider perturbative string decompactification limits. The KK species scale is associated with the higher-dimensional Planck scale \eqref{general:formula:lambda:sp} and, upon identifying the coefficients of the gravitational action with it, in \eqref{second}, we can relate the species scale from KK decompactifications with the string scale, $M_s$:
\be
M_s \simeq \Lambda_\text{sp}\, g_s^{\frac{2}{2+\alpha p}}\,.
    \label{planck:string:relation}
\ee 
We consider $g_s<1$, therefore in general we have $M_s < \Lambda_\text{sp}$. However, we can also consider values $g_s^{\frac{2}{2+\alpha p}}\sim\mathcal{O}(1)$ with $g_s<1$, such that $M_s \sim \Lambda_\text{sp}$.

\medskip

We can also rewrite the species scale in terms of the gravitino mass. Using the Gravitino
Conjecture \eqref{GravConjFinal}
\begin{equation}
R^{-1}=\left(\frac{m_{3/2}}{M_{\text{P}}}\right)^n M_{\text{P}}\,,
\end{equation}
with proportionality constant $\lambda_{3/2}$ to one, then Eq.  \eqref{species:scale:effective}
becomes
\begin{equation}
\label{species:scale:gravitino}
    \Lambda_{\rm sp}
    =
    M_{\rm P}
    \left(
    \frac{m_{3/2}}{M_{\rm P}}
    \right)^{
    \frac{n\alpha p}{2+\alpha p}
    } .
\end{equation}

This expression must be compared with the supersymmetry-breaking scale. In a
quasi-Minkowski vacuum,
\begin{equation}
\label{MSUSY:species:section}
    M_{\rm SUSY}^2
    \sim
    m_{3/2}M_{\rm P},
\end{equation}
and hence
\begin{equation}
\label{MSUSY:explicit:species}
    M_{\rm SUSY}
    =
    M_{\rm P}
    \left(
    \frac{m_{3/2}}{M_{\rm P}}
    \right)^{1/2}.
\end{equation}
Since \(m_{3/2}\ll M_{\rm P}\), the condition
\begin{equation}
    M_{\rm SUSY}\lesssim \Lambda_{\rm sp}
\end{equation}
is equivalent to requiring for the exponent in \eqref{species:scale:gravitino} that
\begin{equation}
\label{species:upper:n}
    n
    \leq
    \frac{2+\alpha p}{2\alpha p}.
\end{equation}
This is the species-scale upper bound on the Gravitino-Conjecture exponent of Eq. \eqref{strong:n:bound}

It is important that, once the bounds on \(n\) are imposed (Section \ref{subsec:bounds:n:p}), \(n\) and \(\alpha\) cannot be
varied independently. The allowed values of \(n\) are correlated with \(\alpha\). At fixed
\(\alpha p\), the Type IIA interval is
\begin{equation}
\label{IIA:n:species:section}
    \frac{2}{\alpha p}
    \leq
    n
    \leq
    \frac{2+\alpha p}{2\alpha p},
    \qquad
    \alpha p\geq 2,
\end{equation}
whereas in Type IIB the lower and upper bounds coincide,
\begin{equation}
\label{IIB:n:species:section}
    n
    =
    \frac{2+\alpha p}{2\alpha p}.
\end{equation}

We can now read off the corresponding range of the species scale. For Type IIB,
substituting \eqref{IIB:n:species:section} into \eqref{species:scale:gravitino} gives
\begin{equation}
\label{IIB:species:MSUSY}
    \Lambda_{\rm sp}
    =
    M_{\rm P}
    \left(
    \frac{m_{3/2}}{M_{\rm P}}
    \right)^{1/2}
    =
    M_{\rm SUSY}.
\end{equation}
Thus, in Type IIB, the species scale saturates the supersymmetry-breaking scale. This saturation is not a universal consequence of the Gravitino Conjecture. It follows
from the perturbative Type IIB scaling used in the main text (section \ref{subsec:bounds:n:p}), which fixes \eqref{IIB:n:species:section}.

For Type IIA, instead, the exponent in \eqref{species:scale:gravitino} ranges between
\begin{equation}
    \frac{2}{2+\alpha p}
    \leq
    \frac{n\alpha p}{2+\alpha p}
    \leq
    \frac12.
\end{equation}
Therefore, at fixed \((p,\alpha)\),
\begin{equation}
\label{IIA:species:range:fixed:alpha}
    M_{\rm SUSY}
    =
    \Lambda_{\rm sp,min}
    \leq
    \Lambda_{\rm sp}
    \leq
    \Lambda_{\rm sp,max}
    =
    M_{\rm P}
    \left(
    \frac{m_{3/2}}{M_{\rm P}}
    \right)^{\frac{2}{2+\alpha p}} .
\end{equation}
If one further allows \(\alpha\) to vary over its full allowed range, the maximal value of
\(\Lambda_{\rm sp}\) is obtained at the largest allowed value of \(\alpha p\). Since
\(\alpha p\leq 6\), one finds
\begin{equation}
\label{IIA:species:range:global}
    M_{\rm SUSY}
    \leq
    \Lambda_{\rm sp}
    \leq
    M_{\rm P}
    \left(
    \frac{m_{3/2}}{M_{\rm P}}
    \right)^{1/4}.
\end{equation}

The crucial point is that the extrema of \(\Lambda_{\rm sp}\) must be taken over the
correlated domain defined by the allowed \((n,\alpha)\) values. Treating \(n\) and
\(\alpha\) as independent variables would incorrectly allow regions in which
\(M_{\rm SUSY}\lesssim \Lambda_{\rm sp}\) is violated.

\bibliographystyle{JHEP}
\bibliography{references-1}

\end{document}